# Flexible Agent-based Modeling Framework to Evaluate Integrated Microtransit and Fixed-route Transit Designs: Mode Choice, Supernetworks, and Fleet Simulation

**Siwei Hu**[1,2]
Email: siweih3@uci.edu
ORCID: 0000-0002-3092-279X

**Michael F. Hyland**[1,2,*]
E-mail: hylandm@uci.edu
ORCiD: 0000-0001-8394-8064

**Ritun Saha**[1,2]
E-mail: rituns@uci.edu

**Jacob J. Berkel**[2,3]
E-mail: berkelj@uci.edu

**Geoffrey Vander Veen**[1,2]
E-mail: vandervg@uci.edu

[*]Corresponding Author
[1]Department of Civil and Environmental Engineering, University of California, Irvine, 92697-3600, CA, USA

[2]Institute of Transportation Studies, University of California, Irvine, 4000 Anteater Instruction and Research Bldg (AIRB), Irvine, 92697-3600, CA, USA

[3]Donald Bren School of Information and Computer Sciences, University of California, Irvine, 6210 Donald Bren Hall, Irvine, 92697-3425, CA, USA

# Abstract

The integration of traditional fixed-route transit (FRT) and more flexible microtransit has been touted as a means of improving mobility and access to opportunity, increasing transit ridership, and promoting environmental sustainability. To help evaluate integrated FRT and microtransit public transit (PT) system (henceforth "integrated fixed-flex PT system") designs, we propose a high-fidelity modeling framework that provides reliable estimates for a wide range of (i) performance metrics and (ii) integrated fixed-flex PT system designs. We formulate the mode choice equilibrium problem as a fixed-point problem wherein microtransit demand is a function of microtransit performance, and microtransit performance depends on microtransit demand. We propose a detailed agent-based simulation modeling framework that includes (i) a binary logit mode choice model (private auto vs. transit), (ii) a supernetwork-based model and pathfinding algorithm for multi-modal transit path choice where the supernetwork includes pedestrian, FRT, and microtransit layers, (iii) a detailed mobility-on-demand fleet simulator called FleetPy to model the supply-demand dynamics of the microtransit service. In this paper, we illustrate the capabilities of the modeling framework by analyzing integrated fixed-flex PT system designs that vary the following design parameters: FRT frequencies and microtransit fleet size, service region structure, virtual stop coverage, and operating hours. We include case studies in downtown San Diego and Lemon Grove, California. The computational results show that the proposed modeling framework converges to a mode choice equilibrium. Moreover, the scenario results imply that introducing a new microtransit service decreases FRT ridership and requires additional subsidies, but it significantly increases job accessibility and slightly reduces total VMT.

**Keywords**: Public Transport, Demand-responsive Transit, Performance Metrics, Mode Choice Equilibrium, Systems Analysis

# Flexible Agent-based Modeling Framework to Evaluate Integrated Microtransit and Fixed-route Transit Designs: Mode Choice, Supernetworks, and Fleet Simulation

Siwei Hu[1,2], Michael F. Hyland[1,2,*], Ritun Saha[1,2], Jacob J. Berkel[2,3], Geoffrey Vander Veen[1,2]

**Abstract**

The integration of traditional fixed-route transit (FRT) and more flexible microtransit has been touted as a means of improving mobility and access to opportunity, increasing transit ridership, and promoting environmental sustainability. To help evaluate integrated FRT and microtransit public transit (PT) system (henceforth "integrated fixed-flex PT system") designs, we propose a high-fidelity modeling framework that provides reliable estimates for a wide range of (i) performance metrics and (ii) integrated fixed-flex PT system designs. We formulate the mode choice equilibrium problem as a fixed-point problem wherein microtransit demand is a function of microtransit performance, and microtransit performance depends on microtransit demand. We propose a detailed agent-based simulation modeling framework that includes (i) a binary logit mode choice model (private auto vs. transit), (ii) a supernetwork-based model and pathfinding algorithm for multi-modal transit path choice where the supernetwork includes pedestrian, FRT, and microtransit layers, (iii) a detailed mobility-on-demand fleet simulator called FleetPy to model the supply-demand dynamics of the microtransit service. In this paper, we illustrate the capabilities of the modeling framework by analyzing integrated fixed-flex PT system designs that vary the following design parameters: FRT frequencies and microtransit fleet size, service region structure, virtual stop coverage, and operating hours. We include case studies in downtown San Diego and Lemon Grove, California. The computational results show that the proposed modeling framework converges to a mode choice equilibrium. Moreover, the scenario results imply that introducing a new microtransit service decreases FRT ridership and requires additional subsidies, but it significantly increases job accessibility and slightly

[*]Corresponding author
  *Email address:* hylandm@uci.edu (Michael F. Hyland)
  [1]Department of Civil and Environmental Engineering, University of California, Irvine, 92697-3600, CA, USA
  [2]Institute of Transportation Studies, University of California, Irvine, 4000 Anteater Instruction and Research Bldg (AIRB), Irvine, 92697-3600, CA, USA
  [3]Donald Bren School of Information and Computer Sciences, University of California, Irvine, 6210 Donald Bren Hall, Irvine, 92697-3425, CA, USA

reduces total VMT.

*Keywords:* Public Transport, Demand-responsive Transit, Performance Metrics, Mode Choice Equilibrium, Systems Analysis

---

**1. Introduction**

In this paper, public transit refers to "the transportation of large numbers of people by means of buses, subway trains, etc., especially within urban areas" (Merriam-Webster, 2024). We include traditional fixed-route transit (FRT) and more flexible microtransit in our definition of public transit. We define microtransit as a publicly owned shared-ride passenger mobility service with flexible vehicle routes and on-demand ride scheduling, where the vehicles are typically 6-10-person vans or small buses.

Prior to the pandemic in the last decade, public transit ridership steadily declined (Erhardt et al., 2022; Lee and Lee, 2022). Researchers have identified several reasons for the decline, including the rise of ride-hailing services (Erhardt et al., 2022), lower gasoline prices (Lee and Lee, 2022), and neighborhood changes in high-density neighborhoods (Lee and Lee, 2022). Figure 1 shows data on public transit ridership from quarter 1 of 1990 to quarter 4 of 2023 in US and Canadian cities, collected by the American Public Transportation Association (Dickens, 2024). Figure 1 shows that bus ridership decreased consistently from 2008 to 2019, while rail mode ridership—including heavy rail, light rail, and commuter rail—experienced a steady increase between 1995 and 2020.

Public transit ridership for all transit modes experienced a steep decline at the onset of the COVID-19 pandemic in March of 2020; since then, many transit agencies have struggled to regain their pre-pandemic ridership. In fact, according to Figure 1, in the fourth quarter of 2023, public transit ridership in major US and Canadian cities only reached 86.07% of the level seen in the fourth quarter of 2019.

Traditional transit systems relying exclusively on FRT networks cannot provide door-to-door or corner-to-corner service. Moreover, in moderate to low population-density areas, where jobs, food, and shopping locations are not spatially clustered, providing good quality service to all transit-dependent travelers with FRT-only systems is nearly impossible. Furthermore, the operation of FRT services in these areas is quite expensive (Guerra and Cervero, 2011). Unlike traditional FRT, microtransit can provide door-to-door or corner-to-corner service to travelers (Ghimire et al., 2024; Hansen et al., 2021; Veve and Chiabaut, 2022).

In recent years, public transit agencies in California have implemented microtransit pilots, including Van Go! from the San Joaquin Regional Transit District in 2019 (San Joaquin Regional Transit District, 2019), Metro Micro from the Los Angeles County Metropolitan Transportation Authority in 2020 (LA Metro, 2020), and SmaRT Ride from the Sacramento Regional Transit District in 2020 (Sacramento Regional Transit District, 2020).

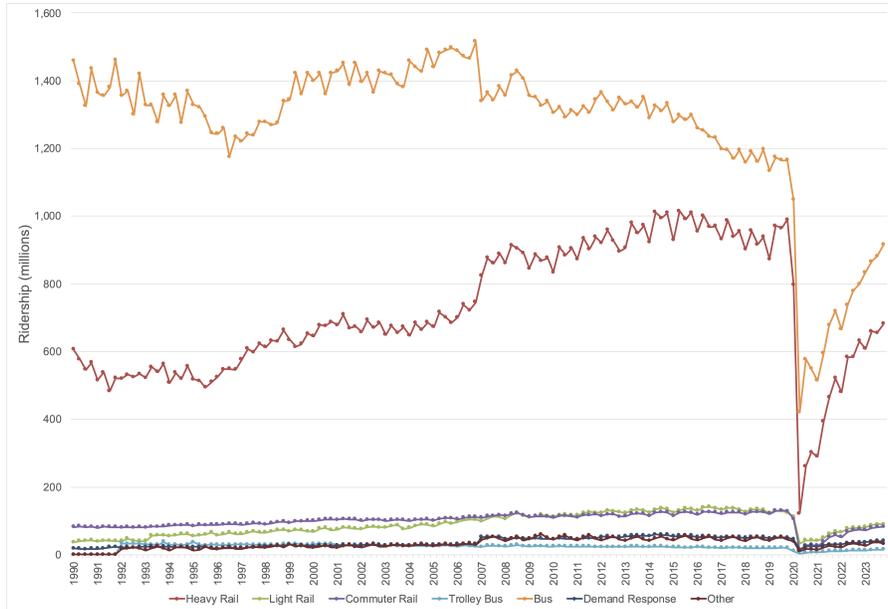

Figure 1: Public ridership from 1990 to 2023 by modes in US and Canadian cities

Due to its flexibility and demand-responsiveness, microtransit is regarded as a complement to traditional FRT, especially in moderate to low population-density areas. Researchers argue that integrating FRT with more flexible microtransit can attract riders to the FRT system (Parks et al., 2020), as well as improve mobility, accessibility, and sustainability of the transportation system (Shaheen and Cohen, 2019).

Figure 2 illustrates an example of an integrated FRT and flexible microtransit system (henceforth, "integrated fixed-flex PT system"). In this diagram, passengers can walk or take microtransit from their origins to FRT stops and walk or take microtransit from FRT stops to their destinations. However, there are few successful real-world examples or helpful guidelines for integrating FRT and microtransit (Currie and Fournier, 2020).

The goal of this study is to develop a high-fidelity modeling framework that researchers and practitioners can use to evaluate alternative integrated fixed-flex PT system designs. We aim to develop a modeling framework with several critical features.

1. Given the flexibility that transit agencies have in terms of designing integrated fixed-flex PT systems and the nature of the underlying design problem, we want the modeling framework to be sensitive to a wide range of FRT and microtransit design parameters.
2. Given that transit agencies have tight budget constraints and address several societal challenges (e.g., accessibility, mobility, environmental impact, and social equity), we want a modeling system capable of providing reliable and useful estimates for a wide range of relevant performance metrics that are sensitive to integrated fixed-flex PT system design parameters.
3. In support of providing reliable estimates, and given the highly heterogeneous sets of travelers in most metropolitan regions (and even within a small area of a city), we want the modeling framework to realistically reflect the heterogeneous sensitivities to travel costs and heterogeneous behavior of travelers in terms of trip mode choice and path choice.
4. Given that scalability is an issue with microtransit services, we want the modeling framework to reflect the substantial impact of demand (overall demand and the spatial and temporal distributions of demand) on the performance of the integrated fixed-flex PT system and, in particular, the microtransit service.

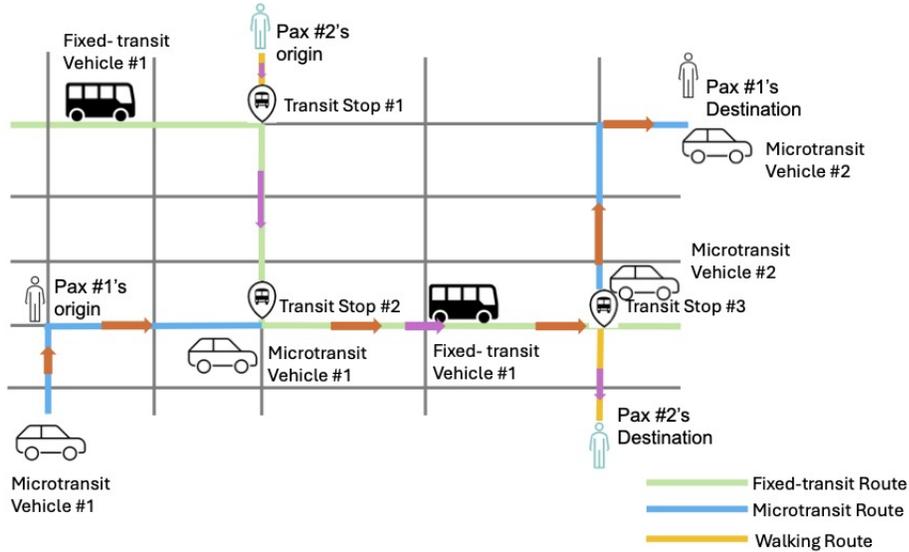

Figure 2: An illustration of FRT and microtransit integration

We found that these features preclude the use of analytical and numerical modeling approaches and require a simulation-based approach. Simulation models (i) require few simplifying assumptions, (ii) can straightforwardly handle a variety of integrated fixed-flex PT system designs, and (iii) can easily produce estimates for a wide variety of system performance metrics. System performance metrics of interest include costs, revenues, ridership, public subsidy, accessibility metrics, mobility metrics, and environmental sustainability metrics. Moreover, given the third model feature, we propose an agent-based modeling framework to capture heterogeneous traveler preferences and behaviors. To address the third model features, we integrate a discrete mode choice model and a supernetwork-based multimodal path choice model. The agents make a discrete mode choice between auto (i.e., drive alone) and composite transit (i.e., the integrated fixed-flex PT system) based on the attributes of these two modes (for the agent's given trip) as well as the agent's attributes. For agents that choose the integrated fixed-flex PT system, we model their discrete path choice using a multi-modal supernetwork and solve a least generalized cost supernetwork path problem for each individual agent subject to their sensitivity to travel costs. We detail the supernetwork model later in the paper, but it is highly generalizable and can handle nearly any set of integrated fixed-flex PT system design parameters. Finally, to meet the fourth model feature, we integrate FleetPy (Engelhardt et al., 2022), an open-source mobility-on-demand (MOD) simulation tool, into the mode and path choice model system. The MOD simulator outputs the performance of the microtransit service as a function of the spatial and temporal demand for microtransit that stems from the mode and path choice models. Given the interplay between microtransit demand and performance, we propose an iterative solution approach to equilibrate microtransit demand and performance.

As far as we know, our proposed modeling framework is the first to capture all of these features. Hence, this paper's primary contribution is the high-fidelity agent- and simulation-based modeling framework developed to evaluate integrated fixed-flex PT system designs. Moreover, while the mode choice model and MOD fleet simulator are standard model components, our proposed multi-modal supernetwork model effectively integrates mode choice, path choice, and microtransit system performance in a manner where microtransit performance and microtransit demand are in equilibrium.

The next section, Section 2, provides background information and reviews related literature. We define the underlying problem in Section 3 and present a solution algorithm and more details on the integrated FRT and microtransit modeling framework in Section 4. Next, through computational experiments, we aim to illustrate the model's capabilities in terms of handling different design parameters (FRT line frequency/headway and microtransit fleet size, service region structure, virtual stops coverage, and operating hours) and producing useful performance metrics. Section 5 describes the model inputs and computational experiments for two case study regions–downtown San Diego and Lemon Grove, a small city in San Diego County. Section 6 displays and discusses the results of the computational experiments. This paper concludes with a summary of the study and future research directions in Section 7.

## 2. Background and Literature Review

As mentioned in Section 1, there are few successful real-world examples of integrated FRT and microtransit systems (Currie and Fournier, 2020). To help inform integrated fixed-flex PT system designs, researchers must first answer the following three questions: (1) What are the critical design variables for an integrated fixed-flex PT system? (2) What goals do transit agencies have that microtransit and integrated fixed-flex PT systems can address, and what metrics should they use to evaluate performance relevant to these goals? (3) How can we properly model integrated fixed-flex PT system designs? This section reviews and summarizes the literature related to these three questions.

In the remainder of this section, we review the studies most relevant to ours. Table 1 summarizes the reviewed literature in this paper in terms of how they address the three questions in the prior paragraph.

Integrated fixed-flex PT system design variables naturally include FRT and microtransit variables. For FRT design variables, most of the existing literature focuses on transit frequency (Calabrò et al., 2023; Luo et al., 2021; Pinto et al., 2020). However, some researchers analyze the spacing between transit lines and stops (Calabrò et al., 2023), start and end stops of regular bus lines (Zhao et al., 2023), and FRT fleet size (Zhao et al., 2023). For microtransit design variables, researchers analyze fleet size (Leffler et al., 2024; Liu et al., 2019; Narayan et al., 2020; Pinto et al., 2020; Zhao et al., 2023), fare policy/structure (Liu et al., 2019; Luo et al., 2021; Narayan et al., 2020), vehicle rebalancing (Luo et al., 2021), and service area size/structure (Zhao et al., 2023).

To illustrate our modeling framework's capabilities, we consider five design variables: FRT frequency and microtransit fleet size, service region structure (continuous vs. contiguous but partitioned), virtual stop coverage, and operating hours. Although not analyzed in this paper, due to a lack of estimated, calibrated, and validated behavioral models, our modeling framework can easily handle a variety of fare structures for integrated fixed-flex PT systems.

For performance metrics, research in the literature collectively consider a wide-variety of metrics. However, each individual study usually only focus on two or three metrics. In the literature, traveler-based metrics include walking time (Calabrò et al., 2023), waiting time (Calabrò et al., 2023; Leffler et al., 2024; Narayan et al., 2020; Pinto et al., 2020), in-vehicle or total travel time (Calabrò et al., 2023; Leffler et al., 2024; Narayan et al., 2020; Zhao et al., 2023) and number of customer rejections (Pinto et al., 2020). Operator-based metrics include capital and operation cost Calabrò et al. (2023) and profit (Liu et al., 2019). Society-based metrics include mode share (Leffler et al., 2024; Narayan et al., 2020) and social welfare (Luo et al., 2021).

To illustrate our modeling framework's capabilities, we focus on four primary performance metrics: subsidy per transit user, 15-minute job accessibility by PT, mode share, and vehicle miles traveled (VMT). Based on conversations with transit agency professionals skeptical of microtransit, the high subsidy per transit trip for microtransit in existing systems is often a point of concern and criticism. On the other hand, for proponents of microtransit, the service's ability to improve access to opportunity over FRT-only service is often cited. Policymakers are often interested in mode share. Moreover, we include VMT as proponents of microtransit often argue for the environmental benefits of people shifting travel from personal vehicle to any transit mode.

For available modes, since the integrated fixed-flex PT system is the focus of this literature review, nearly all studies include FRT and microtransit mode. In addition to FRT and microtransit, some studies include bike mode (Nam et al., 2018; Narayan et al., 2020), some include walking mode (Calabrò et al., 2023; Leffler et al., 2024; Nam et al., 2018; Narayan et al., 2020; Pinto et al., 2020), while others include the auto mode (Narayan et al., 2020). In this paper, we consider four modes: auto, FRT, microtransit, and walking.

Notably, the existing literature uses several different names for what we call "microtransit". Examples include MOD (Luo et al., 2021), fixed-route feeder (FRF) (Calabrò et al., 2023), demand responsive transit (DRT) (Calabrò et al., 2023), adaptive transit scheme (Calabrò et al., 2023), and shared-used autonomous vehicle mobility service (SAMS) in the case where the vehicles are driverless (Pinto et al., 2020).

In addition to design variables, performance metrics, and available modes, it is also crucial for models to capture the critical features of integrated fixed-flex PT system designs. Detailing the critical features mentioned in Section 1, we list five important features that the modeling framework should capture: (1) the heterogeneous behavior of travelers, (2) endogenous demand for the integrated PT system and microtransit in particular, (3) microtransit service fleet dynamics, (4) flexible choice of microtransit service region size (e.g., region-wide or subregion-only) and the service region structure (e.g., one continuous region, multiple contiguous zones, multiple noncontiguous zones), and (5) flexible multimodal path choice that does not require pre-determining modal combinations/sequences or paths themselves. The remainder of the literature review summarizes whether and how the existing studies capture these five important features.

Regarding heterogeneous behavior of travelers, some studies only consider a single class of travelers (Calabrò et al., 2023; Leffler et al., 2024; Liu et al., 2019; Nam et al., 2018; Narayan et al., 2020; Pinto et al., 2020), while others consider two types of travelers (Luo et al., 2021). Different from the existing studies, in this study, we consider each traveler has his/her individual sensitivities to different attributes for transit and auto mode, thus enabling us to model substantial traveler heterogeneity.

For endogenous demand, several studies incorporate mode choice to capture endogenous demand (Leffler et al., 2024; Liu et al., 2019; Luo et al., 2021; Narayan et al., 2020; Pinto et al., 2020), while some studies do not (Calabrò et al., 2023; Nam et al., 2018; Zhao et al., 2023). We use a binary logit mode choice model for auto and composite transit modes to capture endogenous demand for the integrated fixed-flex PT system.

For the microtransit service fleet dynamics, some studies use simulation to capture within-day microtransit fleet dynamics and stochasticity (Leffler et al., 2024; Liu et al., 2019; Pinto et al., 2020), while others do not use a simulator, but formulate the system design problem as a mixed integer linear program (Luo et al., 2021) or nonlinear program (Zhao et al., 2023), or use continuous approximation (CA) (Calabrò et al., 2023). To effectively model the operational performance of the microtransit fleet, we assume that ride requests are unknown a priori, and the fleet operator needs to solve a stochastic dynamic vehicle routing problem. We use the open-source MOD simulator, FleetPy, to model the microtransit service fleet dynamics.

For microtransit service region, in the experimental settings, some studies allow microtransit to only operate in a suburban region (Pinto et al., 2020), while others allow microtransit to operate in multiple but non-contiguous regions (Calabrò et al., 2023; Leffler et al., 2024; Zhao et al., 2023). Only a few studies allow microtransit to operate in the whole region (Liu et al., 2019; Narayan et al., 2020). To effectively capture the optionality related to microtransti service region size and structure that transit agencies have when designing integrated fixed-flex PT systems, we propose a supernetwork based approach where a modeler can readily change the microtransit network layer to capture a variety of service region sizes and structures.

For multimodal path choice, some researchers assume that FRT and microtransit are independent; thus, there are no traveler paths that use both modes (Liu et al., 2019). Others assume that the microtransit service can only serve as an first mile or last mile feeder service for FRT(Calabrò et al., 2023; Luo et al., 2021). Only a few researchers allow traveler paths that include both FRT and microtransit modes, where microtransit legs of a multimodal path are not predefined (Leffler et al., 2024; Nam et al., 2018; Pinto et al., 2020; Zhao et al., 2023).

Even among the studies allow flexible multimodal paths, the approaches are not the same. Nam et al. (2018) adopt a multimodal supernetwork approach that provides the flexibility to integrate different modes into one single network model. However, the system they model fundamentally differs from an integrated fixed-flex PT system with mode choice equilibrium assumptions. Rather, they focus on assigning travelers to the best deterministic (multi-modal) path considering public transit and peer-to-peer ridesharing (i.e., ridesharing without dedicated drivers). Pinto et al. (2020) allow flexible integration of fixed and microtransit paths, but in their computational experiments, the SAMS fleet only operates in a suburban area, Evanston, IL, rather than in the entire Chicago metropolitan area. Pinto et al. (2020) also only allows microtransit-only paths and microtransit to FRT as a first-mile connector. Similar to our approach, Zhao et al. (2023) use a network structure and path-finding algorithm that is fully flexible in terms of modal sequences and inter-modality. However, unlike our study, they assume travelers are homogenous in terms of their sensitivity to travel time and travel cost. Leffler et al. (2024) also permit flexible path integration among FRT, transit, and active mode alternatives. However, it seems that their approach requires multi-modal path enumeration, which is computationally expensive and the computational intensity grows exponentially as the network size expands.

To effectively capture flexible multimodal path choice behavior, we use a multimodal multi-layer supernetwork approach similar to Nam et al. (2018), while considering travelers' heterogeneous sensitivities to link-based modal attributes, such as waiting time, in-vehicle travel time (IVTT), fare, and walk distance in the multimodal pathfinding process. As a result, when calling a least-cost pathfinding algorithm we find the optimal multi-modal path for each individual travel. Even if two travelers share the same origin-destination (OD) pair, their least generalized cost paths could still differ due to different individual preferences.

| Paper (author, year) | Design inputs | Performance metrics | Modes | Model features | | | | |
|---|---|---|---|---|---|---|---|---|
| | | | | Heterogeneous travelers | Endogenous demand | Microtransit dynamics | Microtransit available in the whole region | Flexible multi-modal paths |
| Nam et al. (2018) | N/A | 10-minute accessible area, Bike-sharing mode share, FRT mode share | Bike-sharing, FRT, P2P ridesharing, Walking | | | | | ✓ |
| Liu et al. (2019) | MOD fleet size, MOD service fare | MOD operator's profit | FRT, Microtransit, Ride-hailing, Ridepooling | | ✓ | ✓ | ✓ | |
| Narayan et al. (2020) | FRT to microtransit fare ratio, Microtransit fleet size, | Mode share, Microtransit waiting time, Microtransit travel time, Microtransit utilization | Bike, Car, FRT, Fixed+microtransit, Microtransit, Walking | | ✓ | ✓ | ✓ | |
| Pinto et al. (2020) | FRT frequency, SAMS fleet size | FRT and microtransit boarding rejections, Travelers' waiting time | FRT, FRT+SAMS, SAMS, Walking | | ✓ | ✓ | | ✓ |
| Luo et al. (2021) | FRT frequency, MoD rebalancing flows, Price for each trip | Social welfare | FRT, MoD, MoD+FRT | ✓ | ✓ | | | |
| Calabrò et al. (2023) | FRT Frequency, Spacing between transit lines and stops | Capital and operation costs, IVTT, Walking time, Waiting time | Adaptive transit scheme, MRT, MRT+FRF, Walking | | | | | |
| Zhao et al. (2023) | DRT fleet size, FRT fleet size, Service area of the DRT, Terminal bus stops of regular bus lines | Total fleet size, Total passenger travel time | DRT, FRT | | | | | ✓ |
| Leffler et al. (2024) | Flexible transit fleet size | Ave. exper. IVTT, Ave. exper. waiting time, Mode share | FRT, Microtransit, Walking | | ✓ | ✓ | | ✓ |
| This paper | FRT frequency, Microtransit fleet size, Service region structure, Operating hours, Virtual stop coverage | 15-minute job accessibility, Mode share, Subsidy per transit user, VMT | Auto, FRT, Microtransit, Walking | ✓ | ✓ | ✓ | ✓ | ✓ |

N/A: not applicable; FRT: Fixed-route Transit P2P: Peer-to-peer; MOD: Mobility-On-Demand; SAMS: Shared-used Autonomous vehicle Mobility Service; FMLM: First Mile and Last Mile; MRT: Mass Rapid Transit; FRF: Fixed-route Feeder; DRT: Demand Responsive Transit; Ave.: average; exper.: experienced; VMT: Vehicle Miles Traveled.

Table 1: Related literature

# 3. Problem

This section first provides an overview of the notation used in this paper in Section 3.1. Section 3.2 introduces the assumptions, while Section 3.3 presents the problem statement. Section 3.4 formulates the mode choice equilibrium problem.

## 3.1. Notation

Table 2 summarizes the variables, sets, and coefficients used in this paper.

## 3.2. Assumptions

This paper makes the following assumptions:

1. Following Vansteenwegen et al. (2022)'s taxonomy, we model a microtransit service that is a dynamic-online, many-to-many, fully-flexible, and stop-based service. *Dynamic-online* means the schedule of the microtransit service can change dynamically during operation according to the request received in real time. *Many-to-many* means that passengers can be picked up and dropped off at multiple locations, unlike the many-to-one feeder systems where all travelers share a common destination. *Fully-flexible* means schedules and routes of microtransit services are determined from scratch without any pre-determined patterns. *Stop-based* means passengers must walk to virtual stops to be picked up and dropped off.
2. Arrivals of FRT users at transit stops follow a uniform distribution. As a result, the waiting times of FRT users are half of the route headways. We do not aim to model coordination between FRT and microtransit, as we view this as an operational-level problem rather than a planning and design level problem
3. FRT, auto, and walking networks and network attributes are assumed to be constant, i.e., the level of service of these three modes is independent of the number of travelers using them.
4. Conversely, the microtransit network depends on the demand for microtransit.
5. All modes are available to all travelers.
6. The transit mode and auto mode are independent, i.e., there is no integrated path between auto and transit. This assumption precludes park-and-ride and kiss-and-ride options.

## 3.3. Problem Statement

This subsection describes the underlying problem.
Given:

- a set of travelers with their origins, destinations, departure times, and individual preferences and sensitivities to walking time, waiting time, IVTT, and fare;

| Symbols | Meaning |
| --- | --- |
| Variables | |
| $U_i^m$ | Utility of traveler $i$ selecting mode $m$ |
| $V_i^m$ | Deterministic component of utility of traveler $i$ selecting mode $m$ |
| $\varepsilon_i^m$ | Stochastic component of utility of traveler $i$ selecting mode $m$ |
| $\tilde{w}_{i,T}$ | Walking time of traveler $i$ using transit mode $T$ (min) |
| $w_{i,M}$ | Waiting time of traveler $i$ using microtransit mode $M$ (min) |
| $w_{i,F}$ | Walking time of traveler $i$ using transit mode $F$ (min) |
| $t_{i,M}^{ivt}$ | IVTT of traveler $i$ using microtransit mode $M$ (min) |
| $t_{i,F}^{ivt}$ | IVTT of traveler i using FRT mode $F$ (min) |
| $F_{i,T}$ | Fare of traveler $i$ using transit mode $T$ (microtransit and FRT) ($) |
| $n_{i,F}^t$ | Number of transfers for traveler $i$ using FRT mode $F$ |
| $t_{i,D}^{ivt}$ | IVTT of traveler $i$ using drive alone mode $D$ (min) |
| $c_{i,D}^g$ | Gasoline cost for traveler $i$ using drive alone mode $D$ ($) |
| $Pr_i(m|\mathcal{M})$ | Probability of traveler $i$ choosing mode m from mode choice set $M$ |
| $\Psi_{i,m}^n$ | Traveler $i$'s probability of choosing mode $m$ at iteration $n$ |
| $\eta$ | Perception threshold for mode change |
| $\epsilon$ | Convergence threshold |
| $C_l^F$ | Daily operating cost for line $l$ in the FRT system ($) |
| $T_l$ | Operating hours for line $l$ in the FRT system (hr) |
| $h_l$ | Headway for line $l$ in the FRT system (min) |
| $D_l$ | Line duration for line $l$ in the FRT system (min) |
| $L_l$ | Route total length for route $l$ (mile) |
| $C_F^d$ | Hourly operating cost for one vehicle in FRT system ($) |
| $C_F$ | Daily operating cost for the entire FRT system ($) |
| $C_M$ | Daily operating cost for the entire microtransit system ($) |
| $T_M$ | Daily operating hours for microtransit service ($) |
| $S^f$ | Fleet size for microtransit service |
| $C_M^d$ | Hourly operating cost for microtransit service ($) |
| $VMT_M$ | Microtransit's vehicle miles traveled (mile) |
| $G_M$ | Microtransit's per-mile gasoline cost ($) |
| $\omega$ | Coefficient generation minimum threshold |
| Sets | |
| $R$ | Set of travelers from the synthetic daily demand |
| $R_m$ | Set of travelers choosing mode $m$ |
| $\mathcal{M}$ | Set of the available modes ($\mathcal{M} = \{D, T\}$) |
| Functions | |
| $\boldsymbol{P}_m(R_m)$ | Performance function of mode $m$ |
| $\Psi_m(\boldsymbol{P})$ | Aggregated mode choice probability function for mode $m$ |
| Coefficients | |
| $\beta_{D,0}$ | Alternative specific constraint for auto mode $D$ |
| $\beta_{D,ivt}$ | Coefficient for IVTT for auto mode $D$ |
| $\beta_{D,gas}$ | Coefficient for gasoline cost forcrive alone mode $D$ |
| $\beta_{T,0}$ | Alternative specific constraint for transit mode $T$ |
| $\beta_{T,wk}$ | Coefficient for walking time |
| $\beta_{M,wt}$ | Coefficient for microtransit waiting time |
| $\beta_{F,wt}$ | Coefficient for FRT waiting time |
| $\beta_{M,ivt}$ | Coefficient for microtransit IVTT |
| $\beta_{F,ivt}$ | Coefficient for FRT IVTT |
| $\beta_{F,trfr}$ | Coefficient for FRT transfer penalty |
| $\beta_{T,tr}$ | Coefficient for fare in FRT and microtransit |

Table 2: The notation table

- a multi-modal network with walking, FRT, microtransit, and automobile modes, which are available to all travelers;
- system design inputs – FRT frequency, microtransit fleet size, service region structure, virtual stop coverage, and operating hours;

Find:

1. mode and path choice for every traveler
2. system performance metrics
   - subsidy per transit user,
   - 15-minute job accessibility,
   - mode share,
   - Vehicle Miles Traveled (VMT)

under mode choice equilibrium conditions.

### 3.4. Mode Choice Equilibrium as a Fixed-Point Problem

As Section 2 mentions, the proposed modeling framework solves a mode and path choice problem that ensures equilibrium at the mode choice level. This section presents the mathematical formulation of the mode choice equilibrium problem.

Mode choice equilibrium can be defined as a fixed-point problem (Pinto et al., 2020). Let $R$ be the set of travelers from the synthetic daily travel demand. Let $\mathcal{M}$ be the set of available modes, where $\mathcal{M} = \{T, D\}$, $T$ is the composite transit mode, and $D$ is the drive-alone auto mode. Let $R_m$ be the set of travelers choosing mode $m$ ($m \in \mathcal{M}$). Let $\boldsymbol{P}_m(R_m)$ denote the performance function of mode $m$, where the output is a vector of mode $m$'s performance metrics, with the set of travelers $R_m$ choosing mode m. Let $\Psi_m(\boldsymbol{P})$ be the aggregated mode choice probability function for mode $m$, which takes the performance metrics of all modes, $\boldsymbol{P}$, where $\boldsymbol{P}$ is a vector consisting of performance metrics of the available modes, $\boldsymbol{P} = [\boldsymbol{P}_T(R_T) \quad \boldsymbol{P}_D(R_D)]^T$. Therefore, the set of travelers choosing mode $m$, $R_m$, is given by the choice probability for mode $m$, $\Psi_m(\boldsymbol{P})$, times total number of travelers, $|R|$, in Equation (1):

$$|R_m| = \Psi_m(\boldsymbol{P}) \times |R|, \forall m \in \mathcal{M} \tag{1}$$

Expanding $\boldsymbol{P}$, Equation (1) becomes Equation (2):

$$|R_m| = \Psi_m([\boldsymbol{P}_T(R_T) \quad \boldsymbol{P}_D(R_D)]^T) \times |R|, \forall m \in \mathcal{M} \tag{2}$$

In this study, we assume that the performances (or level of service) of the auto network, $\boldsymbol{P}_D(R_D)$, is constant and does not depend on the sets of travelers in them, $R_D$. So, the solution to the mode choice equilibrium problem, $R_T^*$, satisfies the following fixed-point condition:

$$|R_T^*| = \Psi_T(\boldsymbol{P}_T(R_T^*)|\boldsymbol{P}_D) \times |R| \tag{3}$$

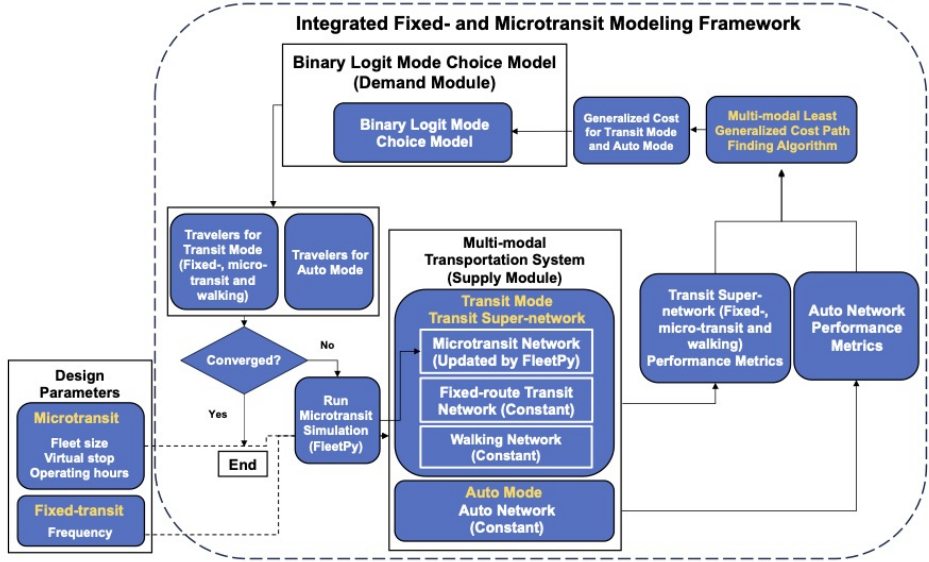

Figure 3: The proposed integrated FRT and microtransit modeling framework

## 4. Solution Algorithm and Modeling Approach

Section 4.1 gives an overview of the proposed solution algorithm to solve the fixed-point mode choice equilibrium problem. Additionally, Section 4.2 delineates the specific models we use for mode choice, supernetwork approach, multimodal path choice, and microtransit fleet simulation. Section 4.3 introduces the perception threshold for mode switching. Section 4.4 presents the convergence criteria for the proposed modeling framework. Lastly, we present performance metrics to evaluate integrated fixed-flex PT system designs in Section 4.5.

### 4.1. Solution Algorithm Overview

Figure 3 gives an overview of the solution algorithm to solve the mode choice equilibrium problem. The algorithm takes design parameters (i.e., fleet size, service region structure, virtual stop coverage, operating hours for microtransit, and frequency for FRT) as input and outputs mode and multi-modal path choices for each traveler, as well as performance metrics, under mode choice equilibrium constraints.

The demand module is shown at the top of Figure 3. We use a binary logit model to capture stochastic traveler mode choice decisions between the auto mode and the integrated fixed-flex PT system. Section 4.2.1 describes the mode choice model in detail. In the solution algorithm, the modal inputs to the binary logit model come from the auto network and integrated system supernetwork model. The demand model outputs the sets of travelers using the integrated system mode and the auto mode.

The supply module is shown at the bottom of Figure 3 and consists of two networks—one supernetwork for the integrated fixed-flex PT system and one network for the drive alone auto mode. Section 4.2.2 details the supernetwork approach, including how it can capture different zonal structures.

As Figure 3 shows, after the demand module outputs travelers for transit and auto modes, the solution algorithm checks whether the system converges (i.e., mode choice equilibrium). If the system has not converged yet, it will run FleetPy again to simulate microtransit service and compute microtransit performance metrics. The system uses microtransit users' waiting time and IVTT to update microtransit waiting and driving links in the transit supernetwork. Once the system converges, it outputs the mode share and the performance metrics associated with each mode.

### 4.2. Model Specifics

#### 4.2.1. Binary logit mode choice model

This section describes the binary logit mode choice model. Let $U_{i,m}$ be the utility of traveler $i$ for mode $m$. The utility consists of a deterministic term, $V_{i,m}$, which could be estimated through observable attributes, and a stochastic term, $\varepsilon_{i,m}$, which accounts for unobservable attributes. Equation (4) displays the utility function:

$$U_{i,m} = V_{i,m} + \varepsilon_{i,m} \tag{4}$$

For transit mode (represented by $T$), the deterministic term, $V_{i,T}$, given by Equation (5):

$$V_{i,T} = \beta_{0,T} + \beta_{1,T}\tilde{w}_{i,T} + \beta_{2,T}w_{i,M} + \beta_{3,T}w_{i,F} + \beta_{4,T}t_{i,M}^{ivt} + \beta_{5,T}t_{i,F}^{ivt} + \beta_{6,T}F_{i,T} + \beta_{7,T}n_{i,F}^{t} \tag{5}$$

where $\tilde{w}_{i,T}$, $w_{i,M}$, $w_{i,F}$, $t_{i,M}^{ivt}$, $t_{i,F}^{ivt}$, $F_{i,T}$, $n_{i,F}^{t}$ are the walking time, microtransit waiting time, FRT waiting time, microtransit IVTT, FRT IVTT, transit fare (FRT and microtransit), and number of transfers for the transit mode, $T$. $\beta_{n,T}$ ($n = 0, \ldots, 7$) represents the coefficients associated with each attribute in the transit utility function.

For the drive alone auto mode (represented by $D$), Equation (6) displays the deterministic term, $V_{i,D}$, :

$$V_{i,D} = \beta_{0,D} + \beta_{1,D}t_{i,D}^{ivt} + \beta_{2,D}c_{i,D}^{g} \tag{6}$$

where $t_{i,D}^{ivt}$, $c_{i,D}^{g}$ are IVTT, and operating cost for traveler $i$ using mode $D$. $\beta_{n,D}$ ($n = 0, 1, 2$) represent the alternative specific constant (ASC), the coefficient for IVTT, and the coefficient for operating cost for mode $D$.

By assuming that the stochastic terms $\varepsilon_{i,m}$ follows an independent and identical Gumbel distribution, one can derive the binary logit mode choice model, where the probability of traveler $i$ choosing mode $m$ is displayed in Equation (7):

$$Pr_i(m|\mathcal{M}) = \frac{e^{V_{i,m}}}{\sum_{m' \in \mathcal{M}} e^{V_{i,m'}}} \tag{7}$$

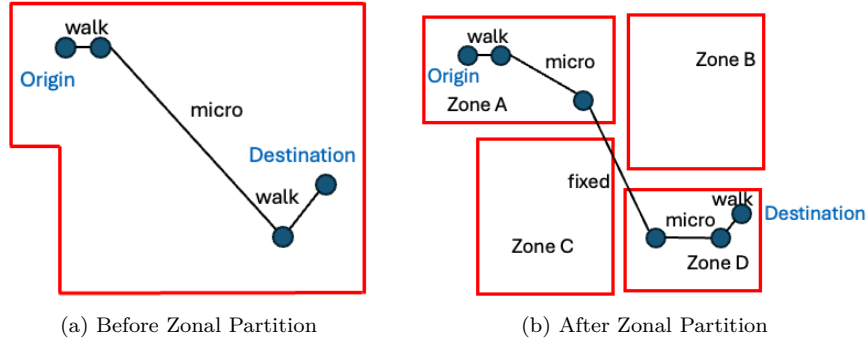

(a) Before Zonal Partition  (b) After Zonal Partition

Figure 4: Example of shortest paths before and after zonal parition

### 4.2.2. Supernetwork approach

To model PT traveler path choice, we use a supernetwork approach, which allows flexible multimodal transit path integration without any pre-defined sequence of modes. The transit supernetwork includes three layers: microtransit, FRT, and walking layers, and connecting links between these layers, representing the waiting times for microtransit and FRT services.

We assume the walking, FRT, and auto networks and network attributes are exogenous. Conversely, we assume the microtransit system performance is a function of the demand for microtransit. We update waiting time links between walking and microtransit layers based on the average waiting time output of FleetPy for a set of microtransit trips. We also update regular (i.e., street segment) link travel times in the microtransit layer based on the average detour travel time output of FleetPy.

With the supernetwork approach, the pathfinding algorithm determines the best path through the network given each modal layer and the connections between the modal layers.

Instead of operating microtransit throughout the entire region, some transit agencies may divide their operating region into different zones, and microtransit can only operate within a certain zone. Our modeling framework readily captures different zonal structures for microtransit services.

To capture the impacts of zonal partitioning, we divide the microtransit layer into 4 different zones. Figure 4 shows the possible best paths from the pathfinding algorithm before and after zonal partition. As Figure 4(a) shows, before zonal partition, the microtransit service can operate throughout the network, allowing the microtransit leg to traverse the entire region. However, when the microtransit service area is partitioned into 4 zones as Figure 4(b) shows, the microtransit can only serve intra-zone trips, while only FRT can serve inter-zone trips.

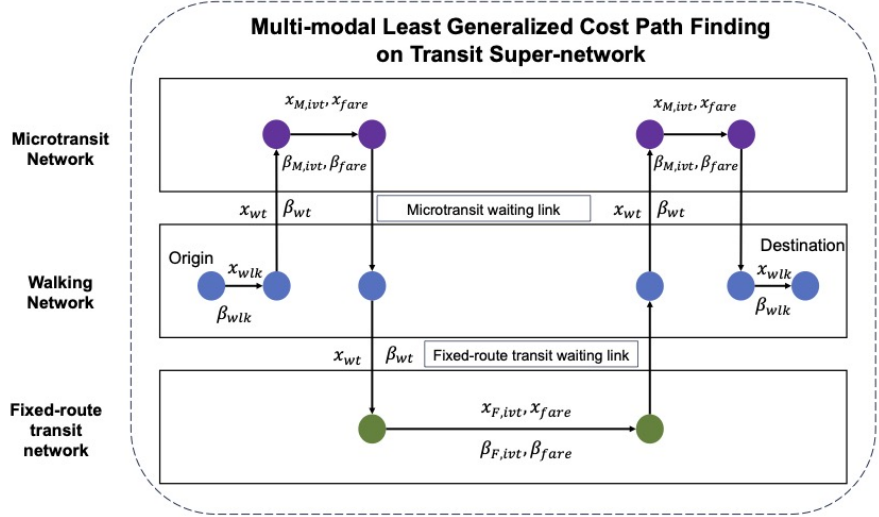

Figure 5: Multi-modal least generalized cost pathfinding process

*4.2.3. Multi-modal least generalized cost pathfinding algorithm*

This paper considers heterogeneous travelers, which means that travelers have different $\beta_{n,T}$'s and $\beta_{n,D}$'s. Hence, if two travelers share the same OD pair, they might not share the same least generalized cost path. Therefore, we calculate the least generalized cost path for each traveler for both transit and auto mode.

We propose a multimodal least generalized cost pathfinding algorithm to calculate the deterministic term for transit mode, $V_{i,T}$, in Equation (5) and the deterministic term for auto mode, $V_{i,D}$, in Equation (6). They are the inputs to the mode choice model in Equation (7).

Figure 5 illustrates the multimodal least generalized cost pathfinding algorithm, which takes both network and travelers' attributes as input. The pathfinding algorithm is modified from Dijkstra's label setting shortest path algorithm (Dijkstra, 1959) considering travelers' attributes ($\beta_{i,m}$'s). The explored nodes are stored in a binary heap structure. The algorithm works as follows: as the pathfinding algorithm traverses each link, it calculates the $\beta_{i,T} \cdot x_{i,T}$ on the current link and sums up the $(\beta_{i,T} \cdot x_{i,T})$'s from a traveler's origin to the current link. When the pathfinding algorithm reaches the traveler's destination, it returns the least generalized cost path for the traveler given his or her OD pair.

*4.2.4. Microtransit service simulation (FleetPy)*

This paper uses FleetPy, a modular open-source MoD service simulator, to model the microtransit service (Engelhardt et al., 2022). FleetPy includes time-and event-driven simulation, vehicle routing, customer demand, infrastructure, and fleet control modules. For the routing module, routes are computed using the Dijkstra algorithm (Engelhardt et al., 2022). The fleet control module uses an immediate response method for passenger-vehicle matching and the algorithm proposed by Alonso-Mora et al. (2017) for repositioning vehicles after they drop off their passengers. Readers can refer to the following reference and the FleetPy GitHub webpage for details (Engelhardt et al., 2022, 2024).

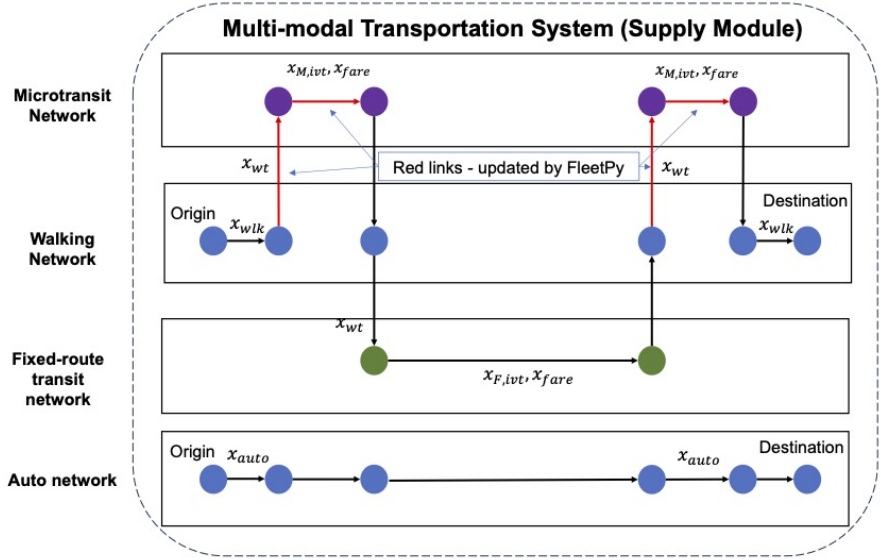

Figure 6: Update microtransit layer with microtransit simulation results

This paper assumes that microtransit vehicles pick up passengers at virtual stops, and passengers walk from their origins to virtual stops and from virtual stops to their destinations. Once a microtransit vehicle picks up a passenger, it drives to the virtual stop that is closest to his or her destination while allowing ride-pooling along the way.

After running the microtransit simulation, the red links in Figure 6 (i.e., microtransit waiting links and traveling links) are updated according to users' waiting times and IVTTs from FleetPy.

### 4.3. Perception Threshold for Mode Change

This paper assumes that travelers have a perception threshold for mode change. The perception threshold is formulated as follows:

1. if the difference of traveler $i$'s probability of choosing mode $T$ in iteration $n$ and iteration $n-1$, $|\Psi_{i,T}^n - \Psi_{i,T}^{n-1}|$, is greater than the mode change threshold $\eta$, namely $|\Psi_{i,T}^n - \Psi_{i,T}^{n-1}| > \eta$, then the model generates a new random number for the binary choice model. A new random number

might lead to a different mode choice result in iteration $n$ than its previous iteration;

2. if $|\Psi_{i,T}^n - \Psi_{i,T}^{n-1}|$, is smaller than or equal to the mode change threshold $\eta$, namely $|\Psi_{i,T}^n - \Psi_{i,T}^{n-1}| \leq \eta$, then traveler $i$ will keep his mode choice in the previous iteration.

This paper sets mode change threshold $\eta$ to be 0.05, which means if the change in traveler's mode choice probability for transit mode or auto mode is less than 0.05, then travelers will remain with the mode chosen in the previous iteration.

### 4.4. Convergence Criterion

This paper uses the squared percentage difference of the mode choice probability between two consecutive iterations of all modes among all travelers as the convergence criterion, as shown in Equation (8):

$$\sum_{i \in I} \sum_{m' \in \mathcal{M}} \frac{(\Psi_{i,m'}^n - \Psi_{i,m'}^{n-1})^2}{\Psi_{i,m'}^{n-1}} \leq \epsilon \tag{8}$$

where $\Psi_{i,m'}^n$ is traveler $i$'s probability of choosing mode $m'$ at iteration $n$ ($n \geq 1$), and $\epsilon$ is the convergence threshold which is set to be 0.01 in this paper. The convergence criterion measures the changes in the mode choice probabilities among all travelers in two consecutive iterations.

### 4.5. Performance Metrics

As mentioned previously, one of the strengths of the proposed modeling framework is the wide range of performance metrics it can reliably estimate. While the modeling framework produces a wide variety of outputs, we focus on four primary performance metrics to compare design alternatives: average subsidy per transit user, jobs accessible within 15 minutes via transit mode, mode share, and VMT.

Section 4.5.1 details how the operating cost, revenue, and subsidy per transit user are calculated, while Section 4.5.2 details the calculation of accessibility. Table 3 shows the units of the performance metrics and Table 4 shows the cost coefficients used in this study.

#### 4.5.1. Cost and revenue for FRT and microtransit

This section presents how the costs of FRT and microtransit are calculated. In terms of FRT, for a single transit line, $l$, the daily operating cost, $C_l^F$, is calculated as in Equation (9):

$$C_l^F = 2 \times \frac{D_l}{h_l} T_l C_F^d + 2 \times \frac{L_l}{h_l} T_l G_F \tag{9}$$

where $T_l, h_l, D_l, L_l$ are operating hours, headway, route duration, and route total length for transit line $l$. The operating hours, $T_l$, is assumed to be 19 hours from 5am to 11:59pm in this paper. $C_F^d$ is the hourly labor cost and $G_F$ is the per-mile gasoline cost for FRT. Number 2 in Equation (9) means that the transit line $l$ operates in two directions. The first term in Equation (9) represents the labor cost and the second term represents the gasoline cost for FRT.

| Performance metrics | Units |
|---|---|
| Operating cost | US Dollar |
| Revenue | US Dollar |
| Subsidy | US Dollar |
| Mode share | % |
| Vehicle miles traveled (VMT) | miles |
| Accessibility (number of jobs within 15 minutes) | count |

Table 3: Performance metrics in this paper

| Coefficient | Meanings | Values |
|---|---|---|
| $T_l$ | Operating hours for FRT line $l$ (hr) | 19 |
| $C_F^d$ | FRT hourly labor cost (\$/hr/vehicle) | 170 |
| $G_F$ | FRT per-mile gasoline cost (\$/mile) | 0.350 |
| $C_M^d$ | Microtransit hourly labor cost (\$/hr/vehicle) | 130 |
| $G_M$ | Microtransit's per-mile gasoline cost (\$/mile) | 0.305 |

Table 4: Values of the cost coefficients

The daily operating cost for the entire FRT system, $C_F$, is calculated as in Equation (10), which sums over all the lines $l$:

$$C_F = \sum_l C_l^F \qquad (10)$$

For FRT, the revenue comes from the \$2.5 flat transit fare, which aligns with the transit fare in the San Diego area (San Diego MTS, 2024).

Microtransit operating daily cost, $C_M$, is calculated as in Equation (11):

$$C_M = T_M S^f C_M^d + VMT_M G_M \qquad (11)$$

where $T_M, S^f, C_M^d$ from the first term are the operating hours, fleet size, and the hourly labor cost for microtransit service, while $VMT_M$ and $G_M$ from the second term are the VMT and per-mile gasoline cost for microtransit service. First term in Equation (11) represents the labor cost and the second term represents the gasoline cost for microtransit service. The revenue in the microtransit service comes from the microtransit fare, which is assumed to be \$1.97/mile in this study.

The performance metric subsidy per transit user is defined as the subsidy for FRT and microtransit divided by the number of FRT and microtransit users.

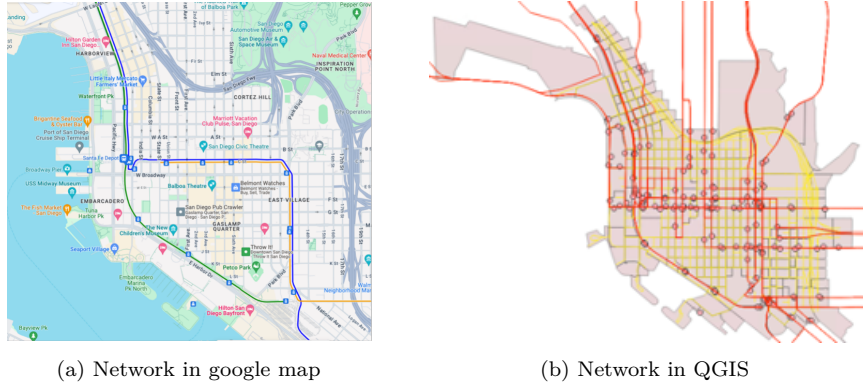

(a) Network in google map  (b) Network in QGIS

Figure 7: Downtown San Diego multi-modal network

*4.5.2. Accessibility*

To measure accessibility, we use the isochrone or cumulative counts approach. Based on the supernetwork at equilibrium and land use information attributed to nodes in the network, for each zone, we count the number of jobs within 15 minutes via the integrated fixed-flex PT system.

## 5. Simulation Setup and Computational Experiments

This section presents the input data and parameters for the computational experiments.

*5.1. Data Input: Network Information*

Figure 7 illustrates the multimodal network in downtown San Diego that includes the auto network and the transit supernetwork. The auto network has 732 nodes and 2,372 directed links, while the FRT layer of the transit supernetwork has 20 transit lines, 162 nodes, and 320 directed links. The transit supernetwork includes walking, microtransit, and FRT layers. The walking layer is modified from the auto network using the walking speed of 2.8mile/hr. The microtransit layer is also modified from the auto network using the auto network travel time times and the microtransit detour ratio, which is the microtransit actual travel time divided by its direct travel time for all microtransit trips. Therefore, the walking, microtransit, and auto networks are three separate networks that have the same topology, but they have different speeds and travel times. This study integrates walking, FRT, and microtransit layers into one transit supernetwork with connecting links (i.e., waiting time for FRT and microtransit) connecting different layers.

Figure 8 illustrates the multimodal network in the city of Lemon Grove, which is a small city with limited FRT coverage in San Diego County. The auto network has 1,099 nodes and 2,816 directed links, while the FRT layer of the transit supernetwork has 6 transit lines, 74 nodes, and 130 directed links.

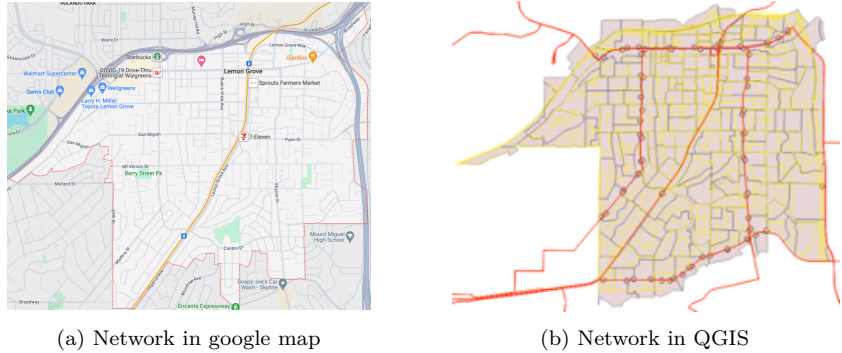

(a) Network in google map  (b) Network in QGIS

Figure 8: Lemon Grove multi-modal network

| From node | To node | Distance (m) | Travel time (sec) | Link type |
|---|---|---|---|---|
| 698 | 3 | 8.698 | 6.953 | 0 (walk) |
| 3 | 698 | 8.698 | 6.953 | 0 (walk) |
| 742 | 743 | 1412 | 120.2 | 1 (FRT) |
| 743 | 742 | 1412 | 120.2 | 1 (FRT) |
| 10 | 742 | 0 | 450 | 2 (FRT waiting) |
| 742 | 10 | 0 | 0 | 2 (FRT waiting) |
| 756 | 772 | 0 | 60 | 3 (FRT transfer) |
| 772 | 756 | 0 | 60 | 3 (FRT transfer) |
| 1471 | 776 | 8.698 | 1.036 | 4 (microtransit) |
| 776 | 1471 | 8.698 | 1.036 | 4 (microtransit) |
| 264 | 1037 | 0 | 795.2 | 5 (microtransit waiting) |
| 1037 | 264 | 0 | 0 | 5 (microtransit waiting) |

Table 5: Supply input: a sample of attributes of the transit supernetwork

The auto network is obtained from the GIS portal of the San Diego Association of Governments (SANDAG) (San Diego Association of Governments, 2024b), while the FRT network is obtained from the General Transit Feed Specification (GTSF) dataset (San Diego Association of Governments, 2024a).

Table 5 shows a sample of the supernetwork with all link types. FRT and microtransit waiting links are virtual links with zero link lengths. Waiting time links for boarding directions have non-zero travel times, while those for alighting directions have zero travel times.

### 5.2. Data Input: Demand Information

In this study, we use synthetic travel demand generated from ActivitySim (Association of Metropolitan Planning Organizations, 2024), an open-source activity-based travel demand modeling platform. ActivitySim outputs the travel demand for a weekday. ActivitySim outputs 46,241 trips for Downtown San

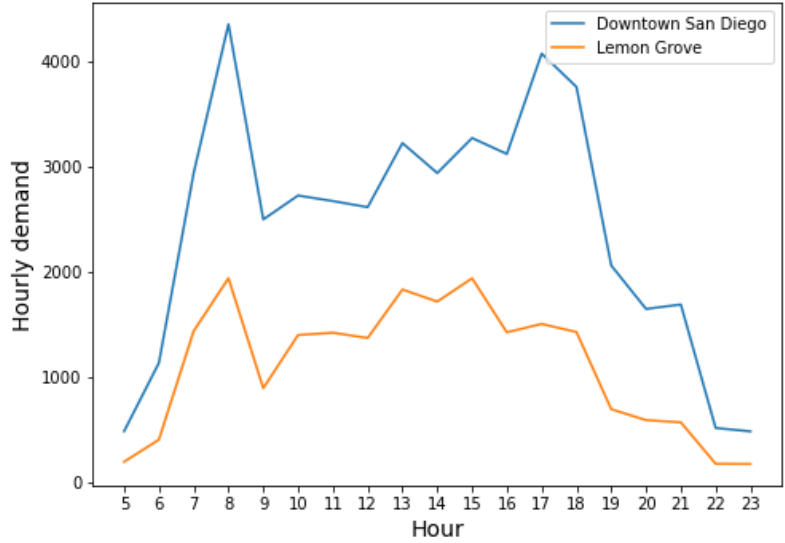

Figure 9: Temporal demand patterns for downtown San Diego and Lemon Grove

| dp_time | O | D | rq_id | $\beta_{c,0}$ | $\beta_{c,ivt}$ | $\beta_{c,gas}$ | $\beta_{T,0}$ | $\beta_{T,wk}$ | $\beta_{T,wt}$ | $\beta_{m,ivt}$ | $\beta_{f,ivt}$ | $\beta_{f,trfr}$ | $\beta_{T,fr}$ |
|---|---|---|---|---|---|---|---|---|---|---|---|---|---|
| 18006 | 384 | 498 | 0 | 0 | 0.129 | 0.631 | 0.456 | 0.018 | 0.089 | 0.161 | 0.115 | 0.469 | 0.778 |
| 18011 | 186 | 297 | 1 | 0 | 0.197 | 0.739 | 0.395 | 0.038 | 0.114 | 0.103 | 0.160 | 0.502 | 0.947 |
| 18014 | 423 | 161 | 2 | 0 | 0.085 | 0.622 | 0.426 | 0.195 | 0.101 | 0.131 | 0.088 | 0.519 | 0.697 |

Table 6: Demand input: a sample of travelers' demand profiles

Diego from 5am to 11:59pm and 21,074 trips for the city of Lemon Grove. Figure 9 illustrates the temporal demand pattern for downtown San Diego and Lemon Grove, where there is a morning peak from 7-9am and an evening peak from 6-7pm. Since the first trip leaves at 5 am in ActivitySim, the horizontal axis in Figure 9 starts at 5am. Figure 10 shows the spatial demand patterns (the spatial distribution of trips' origins) for downtown San Diego and Lemon Grove. Table 6 shows a sample of travelers' demand profiles, including their request IDs, origins, destinations, departure times, and their individual sensitivities to transit fare, fuel costs, in-vehicle travel time, walk time, wait time, and transfers.

*5.3. Simulation Settings and Coefficients*

In the main set of computational experiments, we vary four design parameters for downtown San Diego and Lemon Grove. Below is a list of design parameters and the values that vary across the computational experiments.

1. FRT headway (15 and 30 mins)
2. Virtual stops coverage (75% virtual stops coverage and 100% virtual stops coverage)

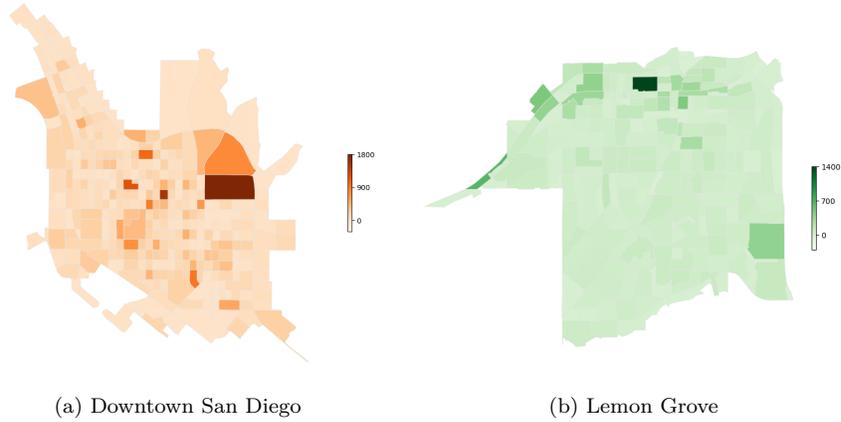

(a) Downtown San Diego    (b) Lemon Grove

Figure 10: Spatial demand patterns (origins)

3. Fleet size (10, 15, and 20 vehicles)
4. Operating periods
   - AM (5-10am) and PM (3-8pm) (operating hours: 10hr)
   - AM (5-10am), MD (10am-3pm), and PM (3-8pm) (operating hours: 15hr)

In addition to the 24 scenarios above, we also analyze two FRT-only scenarios (headway 15 minutes and 30 minutes) and 12 micro transit-only scenarios, resulting in 38 scenarios in total.

For FRT, without loss of generality and to limit the number of computational experiments, we set all transit lines to a headway of either 15 minutes or 30 minutes.

For virtual stop coverage, we analyze two scenarios, 75% and 100%. In the 100% case, every node in the walking network, and therefore every trip origin and destination, are eligible for pickup and drop-off. In the 75% case, only 75% of nodes in the walking network are eligible for pickup and drop-off, meaning that some travelers must walk to a pickup node from their trip origin or from a drop-off node to their final trip destination.

For microtransit fleet size, we analyze three scenarios: 10, 15, and 20 vehicles. We use these small fleet size values as most microtransit services in operation only include 10-20 vehicles.

For operating periods, we study two scenarios: AM and PM (10 operating hours) and AM, MD, and PM (15 operating hours). Table 7 shows all testing scenarios.

For the individual coefficients in the binary mode choice model ($\beta$'s in Table 6), this paper adopts the coefficients of the estimated multinomial logit mode choice model in Frei et al. (2017) who estimated the model for FRT, microtransit, and drive alone in the Chicago region. In this study, we modify the coefficients to match the commute mode share in downtown San Diego and Lemon Grove

| Scenario ID | Transit Mode | Headway (min) | Virtual stop (%) | Fleet size | Operating Periods |
|---|---|---|---|---|---|
| 0 | Micro only | 0 | 75 | 10 | ['AM', 'PM'] |
| 1 | Micro only | 0 | 75 | 10 | ['AM', 'MD', 'PM'] |
| 2 | Micro only | 0 | 75 | 15 | ['AM', 'PM'] |
| 3 | Micro only | 0 | 75 | 15 | ['AM', 'MD', 'PM'] |
| 4 | Micro only | 0 | 75 | 20 | ['AM', 'PM'] |
| 5 | Micro only | 0 | 75 | 20 | ['AM', 'MD', 'PM'] |
| 6 | Micro only | 0 | 100 | 10 | ['AM', 'PM'] |
| 7 | Micro only | 0 | 100 | 10 | ['AM', 'MD', 'PM'] |
| 8 | Micro only | 0 | 100 | 15 | ['AM', 'PM'] |
| 9 | Micro only | 0 | 100 | 15 | ['AM', 'MD', 'PM'] |
| 10 | Micro only | 0 | 100 | 20 | ['AM', 'PM'] |
| 11 | Micro only | 0 | 100 | 20 | ['AM', 'MD', 'PM'] |
| 12 | Fixed only | 15 | 0 | 0 | 0 |
| 13 | Micro+Fixed | 15 | 75 | 10 | ['AM', 'PM'] |
| 14 | Micro+Fixed | 15 | 75 | 10 | ['AM', 'MD', 'PM'] |
| 15 | Micro+Fixed | 15 | 75 | 15 | ['AM', 'PM'] |
| 16 | Micro+Fixed | 15 | 75 | 15 | ['AM', 'MD', 'PM'] |
| 17 | Micro+Fixed | 15 | 75 | 20 | ['AM', 'PM'] |
| 18 | Micro+Fixed | 15 | 75 | 20 | ['AM', 'MD', 'PM'] |
| 19 | Micro+Fixed | 15 | 100 | 10 | ['AM', 'PM'] |
| 20 | Micro+Fixed | 15 | 100 | 10 | ['AM', 'MD', 'PM'] |
| 21 | Micro+Fixed | 15 | 100 | 15 | ['AM', 'PM'] |
| 22 | Micro+Fixed | 15 | 100 | 15 | ['AM', 'MD', 'PM'] |
| 23 | Micro+Fixed | 15 | 100 | 20 | ['AM', 'PM'] |
| 24 | Micro+Fixed | 15 | 100 | 20 | ['AM', 'MD', 'PM'] |
| 25 | Fixed only | 30 | 0 | 0 | 0 |
| 26 | Micro+Fixed | 30 | 75 | 10 | ['AM', 'PM'] |
| 27 | Micro+Fixed | 30 | 75 | 10 | ['AM', 'MD', 'PM'] |
| 28 | Micro+Fixed | 30 | 75 | 15 | ['AM', 'PM'] |
| 29 | Micro+Fixed | 30 | 75 | 15 | ['AM', 'MD', 'PM'] |
| 30 | Micro+Fixed | 30 | 75 | 20 | ['AM', 'PM'] |
| 31 | Micro+Fixed | 30 | 75 | 20 | ['AM', 'MD', 'PM'] |
| 32 | Micro+Fixed | 30 | 100 | 10 | ['AM', 'PM'] |
| 33 | Micro+Fixed | 30 | 100 | 10 | ['AM', 'MD', 'PM'] |
| 34 | Micro+Fixed | 30 | 100 | 15 | ['AM', 'PM'] |
| 35 | Micro+Fixed | 30 | 100 | 15 | ['AM', 'MD', 'PM'] |
| 36 | Micro+Fixed | 30 | 100 | 20 | ['AM', 'PM'] |
| 37 | Micro+Fixed | 30 | 100 | 20 | ['AM', 'MD', 'PM'] |

Table 7: Testing scenarios

| Coefficients | Mean (SD) | Stand_Dev (SD) | Mean (LG) | Stand_Dev (LG) | Threshold |
|---|---|---|---|---|---|
| $\beta_{D,0}$ | 0 | 0 | 0 | 0 | 0 |
| $\beta_{D,ivt}$ | 0.184 | 0.047 | 0.198 | 0.047 | 0.01 |
| $\beta_{D,gas}$ | 0.994 | 0.377 | 0.579 | 0.377 | 0.05 |
| $\beta_{T,0}$ | 0.022 | 0.04 | 0.292 | 0.04 | 0 |
| $\beta_{T,wk}$ | 0.213 | 0.140 | 0.329 | 0.140 | 0.01 |
| $\beta_{M,wt}$ | 0.104 | 0.022 | 0.094 | 0.022 | 0.01 |
| $\beta_{F,wt}$ | 0.069 | 0.022 | 0.082 | 0.022 | 0.01 |
| $\beta_{M,ivt}$ | 0.104 | 0.022 | 0.104 | 0.022 | 0.01 |
| $\beta_{F,ivt}$ | 0.102 | 0.029 | 0.106 | 0.029 | 0.01 |
| $\beta_{F,trfr}$ | 0.504 | 0.022 | 0.504 | 0.022 | 0.01 |
| $\beta_{T,tr}$ | 0.554 | 0.377 | 0.554 | 0.377 | 0.05 |

SD: Downtown San Diego; Stand_Dev: Standard Deviation; LG: Lemon Grove.

Table 8: Coefficients used to generate $\beta$'s in this paper

according to the American Community Survey (U.S. Census Bureau, 2024). We generate individual $\beta$ values based on the normal distribution and the mean and standard deviation in Table 6. Since each $\beta$ should be nonnegative, we truncate the normal distribution at a lower bound $\omega$ (either 0, 0.01 or 0.05) in the generation process to make sure all the generated $\beta$'s are positive (i.e., $\beta = \max(\tilde{\beta}, \omega)$). Table 8 presents the mean, standard deviation, and threshold for each coefficient in San Diego and Lemon Grove.

# 6. Results

This section presents the results of the computational experiments. We first illustrate that the model system converges - i.e., the microtransit system performance and demand equilibrate - in Section 6.1. In Section 6.2, we present the results of the main computational experiments. We display four key performance metrics, namely, transit subsidy per user, number of jobs accessible within 15 minutes, mode share, and VMT for the different scenarios in Section 6.2.1 through Section 6.2.3. We present additional model outputs for zonal accessibility, individual trip length distribution, and transit line usage in Section 6.3. Finally, in 6.4, we analyze the impact of dividing the microtransit service region into four zones where microtransit vehicles cannot transport travelers between zones.

## 6.1. Convergence Analysis

Figure 11 shows the convergence patterns for downtown San Diego and Lemon Grove across the service design variants. The convergence gap in the figure is the squared percentage difference in Equation (8). The results indicate that the proposed solution algorithm converges under all scenarios in both study areas. For the downtown San Diego network, all scenarios converge within 7 iterations, while for the Lemon Grove network, all scenarios converge within 4 iterations.

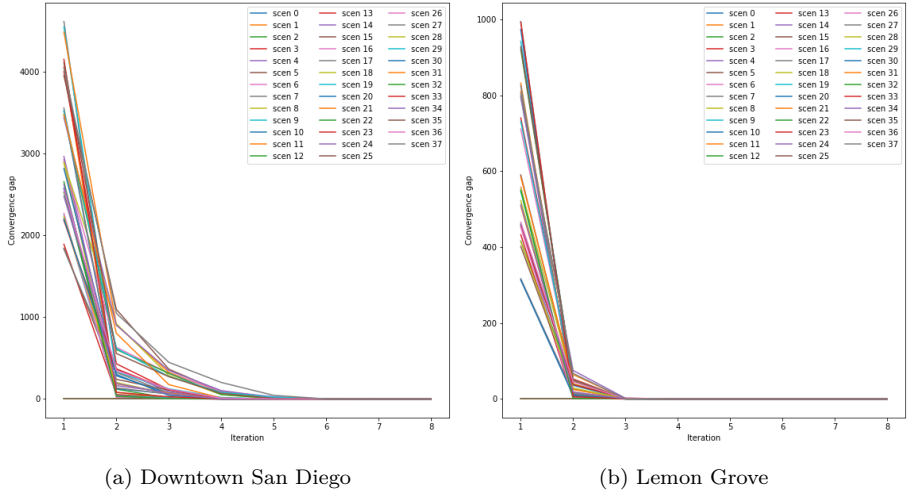

(a) Downtown San Diego　　　　(b) Lemon Grove

Figure 11: Convergence patterns for downtown San Diego (left) and Lemon Grove (right) among 38 scenarios

## 6.2. Main Computational Experiments

### 6.2.1. Impacts of fleet sizes

Figure 12 illustrates the impacts of fleet sizes and transit frequencies on performance metrics in downtown San Diego, while Figure 13 illustrates their impacts on Lemon Grove. The results indicate that as fleet size increases in both study areas, the subsidy per transit user increases, job accessibility increases, car mode share decreases, and total VMT increases. Additionally, as headway increases (from 15 minutes to 30 minutes to "no transit service"), the subsidy per transit user increases, job accessibility decreases, auto mode share increases, and VMT is nonlinear.

The fact that subsidy per transit user increases with microtransit fleet size means that the cost of adding more microtransit vehicles outweighs the additional revenue. The subsidy per transit user is defined in Section 4.5.1. Figure 12(a) shows that for downtown San Diego, when microtransit service is available (fleet size ≥ zero), the 15-minute headway design requires the lowest subsidy per transit user compared to no FRT (microtransit only) and 30-minute transit headway. For downtown San Diego, when microtransit service is unavailable, the subsidy per transit user for a 30-minute headway is negative, which means that the FRT service makes a profit under this scenario. This is because FRT is popular in downtown San Diego and the reason is quite small, so the cost of operating the transit lines in just this area is relatively low. So given the high transit mode share in the region, it is possible that the revenue from the transit fare is more than the operating cost, making the FRT system profitable. But it is not the case in Lemon Grove, where FRT is much less popular. It is also worth noting that Lemon Grove requires a higher subsidy per transit user because fewer transit users in Lemon Grove generate less revenue. This difference implies that opening a microtransit service in a low transit coverage area, such as Lemon Grove, requires a higher subsidy per user compared to a high transit coverage area.

Figure 12(b) and Figure 13(b) indicate that as the microtransit fleet sizes increase, more jobs will be available within 15 minutes. When microtransit service is not available (fleet size = 0), downtown San Diego has more accessible jobs within 15 minutes than Lemon Grove because it has a higher coverage FRT network. When microtransit service becomes available, the number of accessible jobs in Lemon Grove is tripled compared to the scenario without microtransit service. This difference implies that in the low transit coverage area, opening microtransit service can provide a significant increase in accessibility.

Figure 12(c) and Figure 13(c) show that as the microtransit fleet sizes increase, its mode share will increase, while the mode share for FRT, walking, and car mode will decrease. This is because microtransit competes with the three other modes. Microtransit mode share ranges from 6% to 12% in different testing scenarios in Lemon Grove, while it ranges from 3% to 9% in downtown San Diego. This difference implies that the microtransit service will be more popular in areas of low transit coverage than in areas of high transit coverage.

Figure 12(d) and 13(d) show that as the vehicle fleet sizes increase, the total vehicle miles traveled (VMT) remains at the same level, while the car mode VMT decreases. This implies that opening microtransit service will not reduce the total VMT, but will decrease the auto-mode VMT. So, for sustainability and greenhouse gas reduction purposes, opening a microtransit service will not be a good solution.

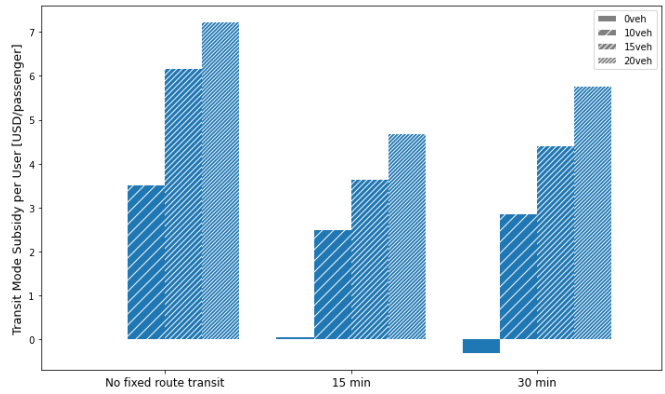
(a) subsidy per transit user

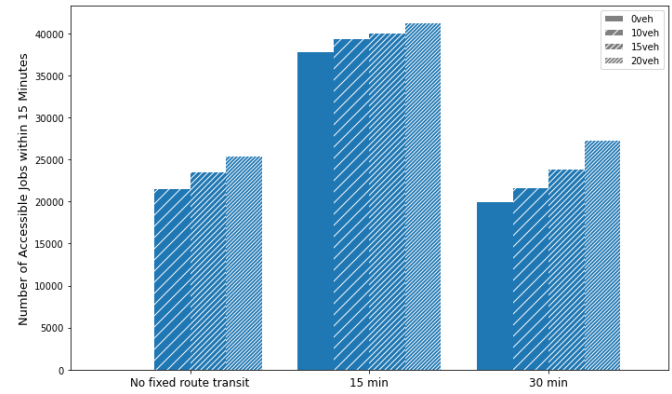
(b) 15-minute job accessibility

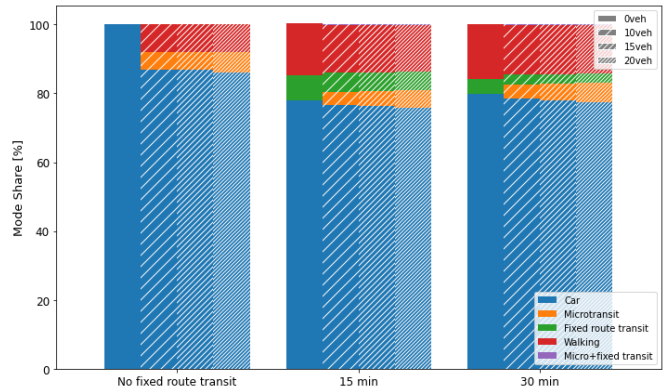
(c) mode share

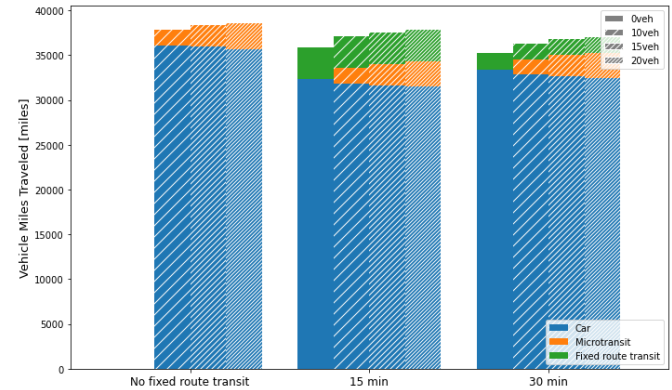
(d) vehicle miles traveled

Figure 12: Impacts of fleet size and transit frequencies on downtown San Diego network (virtual stop=75%, operating period=[AM and PM])

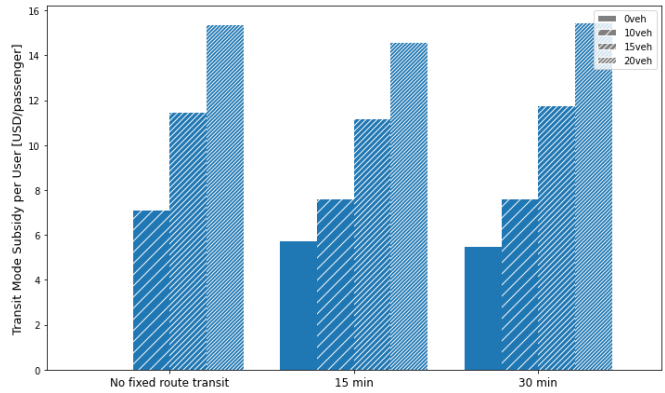

(a) subsidy per transit user

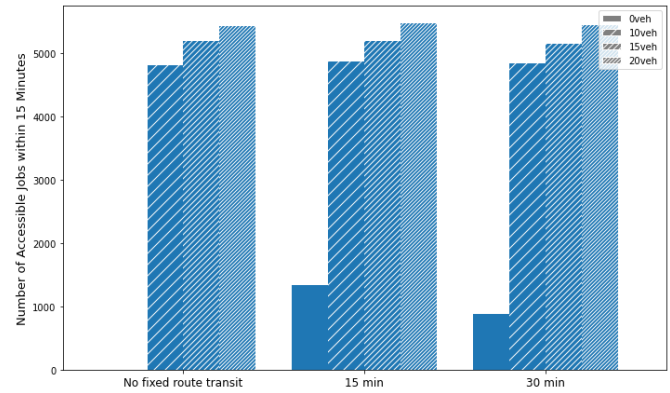

(b) 15-minute job accessibility

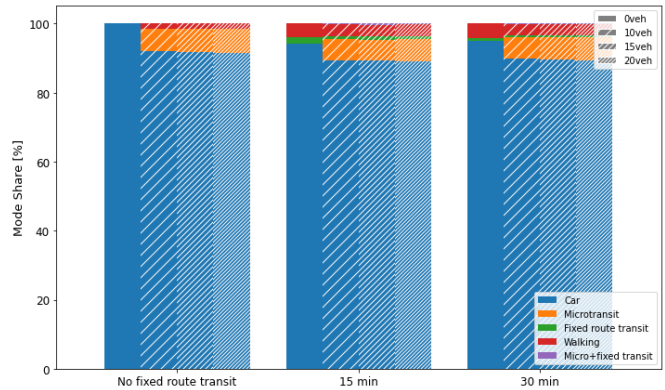

(c) mode share

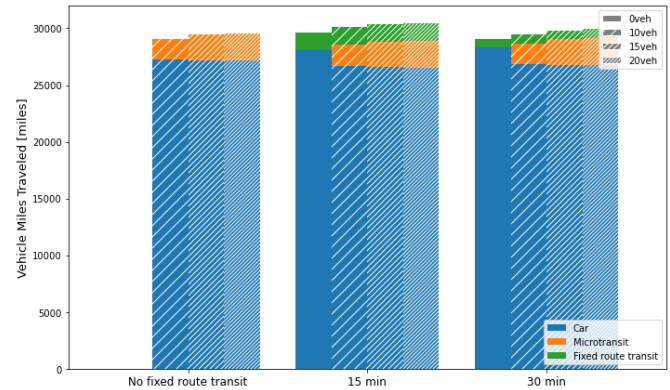

(d) vehicle miles traveled

Figure 13: Impacts of fleet size and transit frequencies on Lemon Grove network (virtual stop=75%, operating period=[AM and PM])

*6.2.2. Impacts of virtual stop coverage*

Figure 14 illustrates the impacts of different virtual stop coverage and transit frequencies on downtown San Diego, while Figure 15 illustrates their impacts on Lemon Grove. Figure 14(a) and Figure 15(a) show that as the coverage of the virtual stop increases from 75% to 100%, both downtown San Diego and Lemon Grove require a lower subsidy per transit user. This is because increasing virtual stop coverage will generate more revenue for microtransit service while maintaining the operating cost. The increase in microtransit revenue results from the increase in microtransit users, because a higher virtual stop coverage leads to fewer walking distances for passengers. This result implies that to keep the subsidy per transit user low, agencies should always set virtual stops at 100%, which means that microtransit vehicles can pick up passengers anywhere.

Figure 14(b) and Figure 15(b) show that the increase in virtual stop coverage does not increase the number of accessible jobs within 15 minutes in downtown San Diego, but it increases the accessibility in Lemon Grove where the FRT coverage is low.

Figure 14(c) and Figure 15(c) show that the increase in the virtual stop coverage increase the microtransit mode share. For areas with lower FRT coverage the increase in virtual stop coverage will lead to an even higher increase in microtransit mode share, as was the case in Lemon Grove.

Figure 14(d) and Figure 15(d) show that increasing virtual stop coverage can lead to a small increase in microtransit VMT, but it won't necessarily increase the total VMT of all modes.

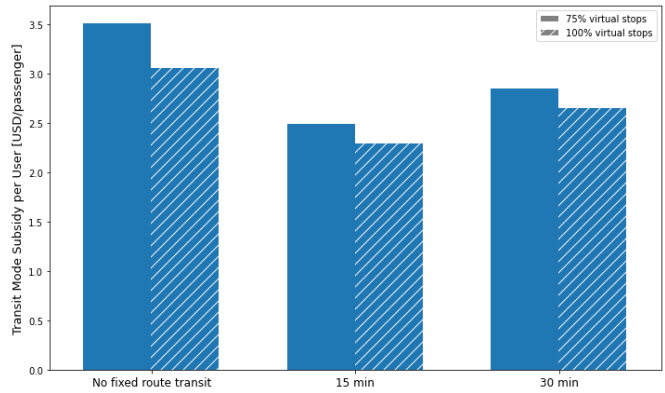
(a) subsidy per transit user

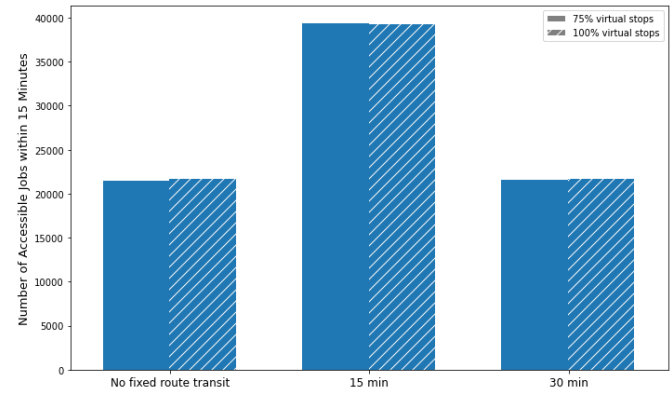
(b) 15-minute job accessibility

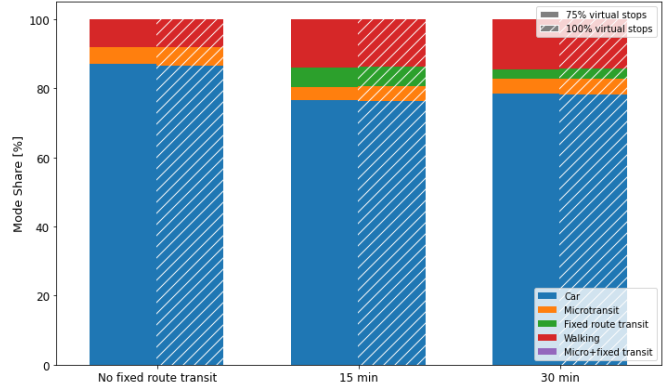
(c) mode share

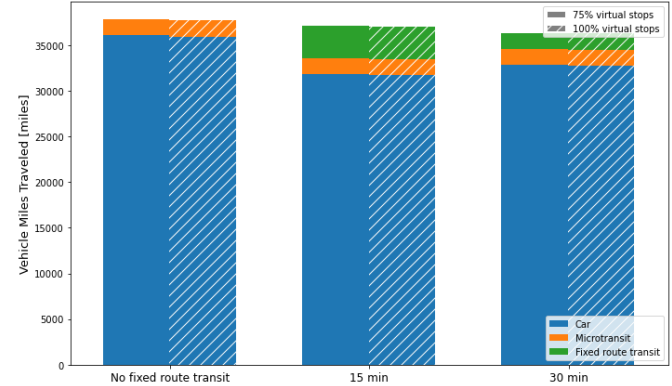
(d) vehicle miles traveled

Figure 14: Impacts of virtual stop coverage and transit frequencies on downtown San Diego network (fleet size=10 veh, operating period=[AM and PM])

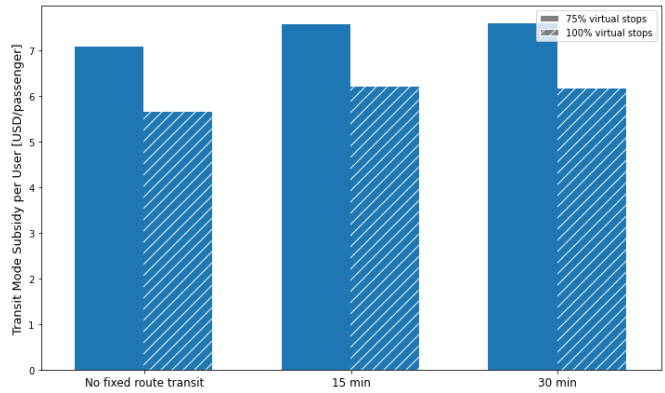
(a) subsidy per transit user

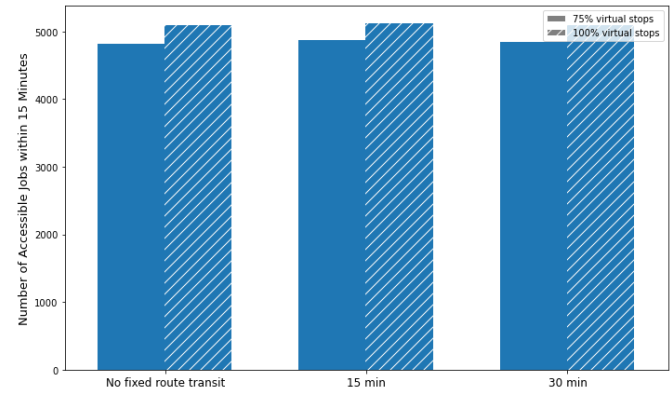
(b) 15-minute job accessibility

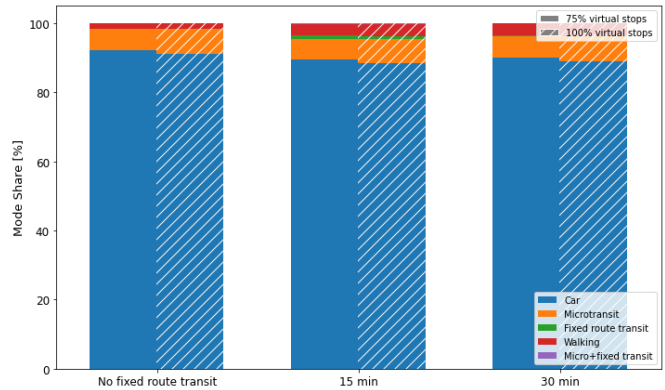
(c) mode share

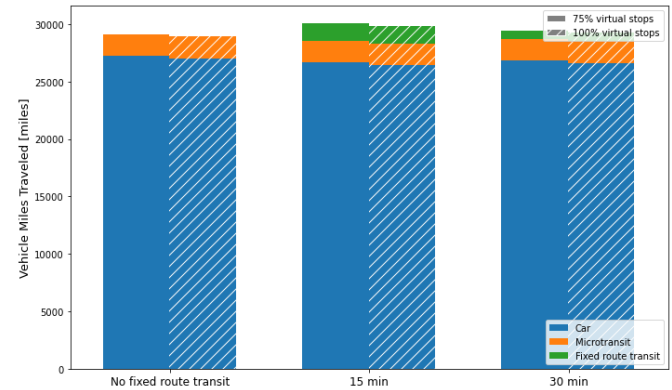
(d) vehicle miles traveled

Figure 15: Impacts of virtual stop coverage and transit frequencies on Lemon Grove network (fleet size=10 veh, operating period=[AM and PM]

### 6.2.3. Impacts of operating periods

Figure 16 illustrates the impacts of different operating periods and transit frequencies for downtown San Diego, while Figure 17 illustrates their impacts for Lemon Grove. Figure 16(a) and Figure 17(a) show that as operating periods increase from [AM, PM] (10 hours) to [AM, MD and PM] (15 hours), the subsidy per transit user increases in downtown San Diego, but decreases in Lemon Grove. The results imply that given the relatively low demand, opening a microtransit service during the MD period (10am-3pm) makes the service more costly in San Diego, while it makes the service less costly in Lemon Grove. This is because Lemon Grove has a low-coverage transit network. The extension of microtransit operating periods encourages more users to use the microtransit service, generating more revenue. However, due to the convenient transit network in downtown San Diego and the low demand of the MD period, opening the microtransit service during this period requires more subsidies in downtown San Diego. The difference implies that for high transit coverage areas such as downtown San Diego, it is better to keep microtransit service operating only during the morning and evening peak periods, but for low transit coverage areas such as Lemon Grove, it is better to include the MD period for the microtransit service operation. But Lemon Grove still requires a higher subsidy per transit user than downtown San Diego (see y axes in Figures 16(a) and 17(a)).

Figure 16(b) and Figure 17(b) show that the 15-minute job accessibilities within the operating period remain stable as the operating periods increase from [AM, PM] (10 hours) to [AM, MD and PM] (15 hours). Figure 16(c) and Figure 17(c) show that as operating periods increase, the microtransit mode share increases in both downtown San Diego and Lemon Grove, but Lemon Grove has an even higher increase. However, the share of FRT modes decreases as operating periods increase. For microtransit-only scenarios (no FRT), increasing operating periods not only increases the microtransit mode share, but also increases the walking mode share, because the car mode users are shifted to microtransit mode, which includes walking legs, and thus increases the walking mode share.

Figure 16(d) and Figure 17(d) show that increasing operating periods leads to an increase in microtransit VMT. However, total VMT decreases as operating periods increase under microtransit-only scenarios (no FRT), but increases in microtransit and FRT scenarios (15-minute and 30-minute headways). However, such a change in VMT is not significant, which implies that opening microtransit service or prolonging microtransit service hours will not be an effective solution for environmental sustainability.

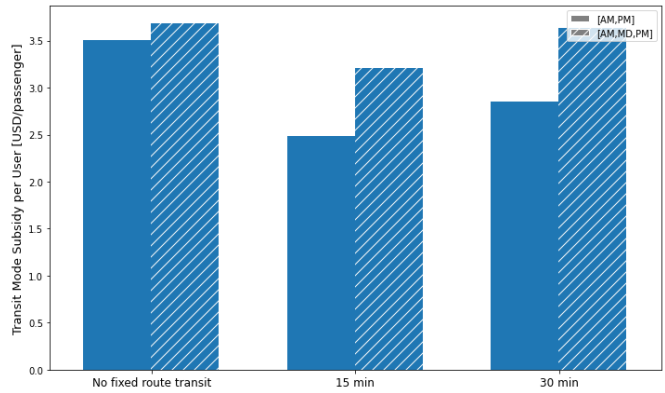
(a) subsidy per transit user

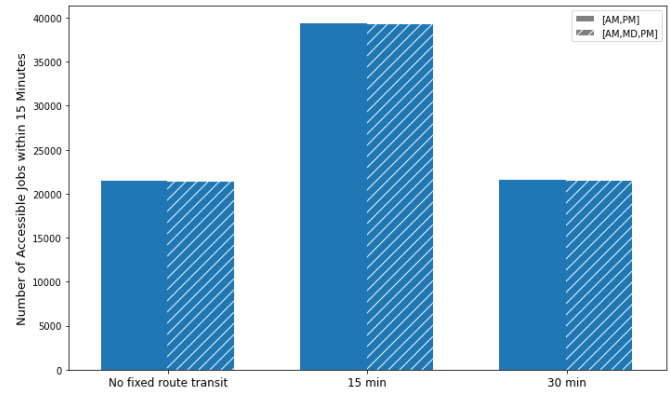
(b) 15-minute job accessibility

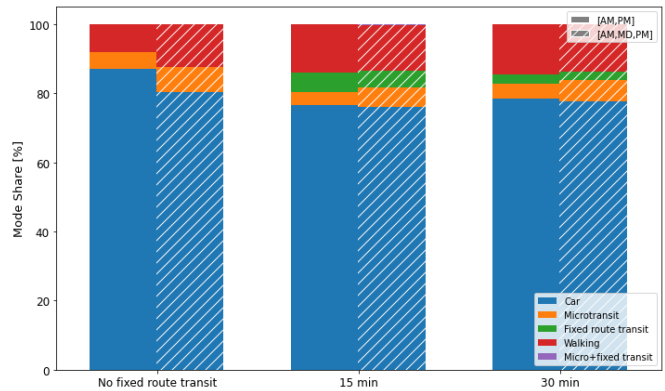
(c) mode share

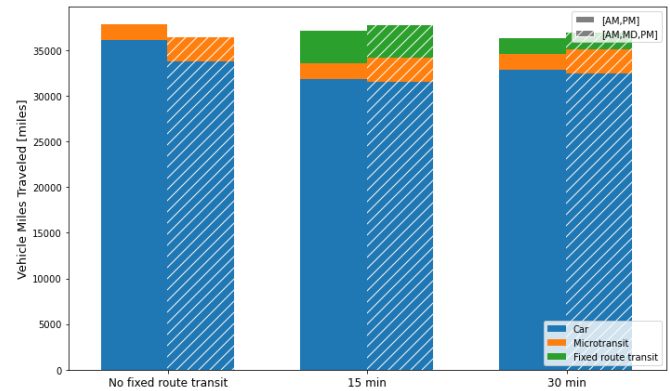
(d) vehicle miles traveled

Figure 16: Impacts of operating periods and transit frequencies for downtown San Diego network (virtual stop=75%, fleet size=10 vehicles)

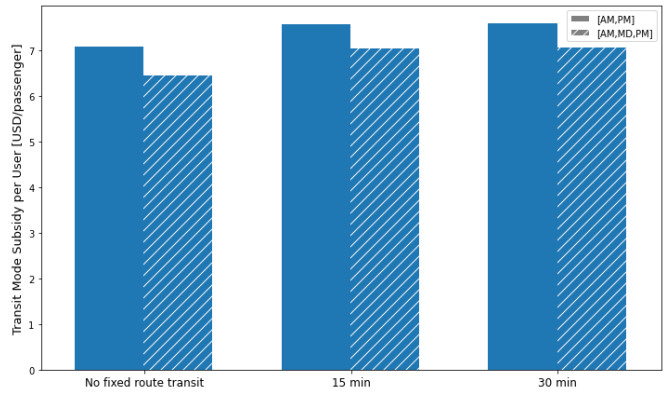
(a) subsidy per transit user

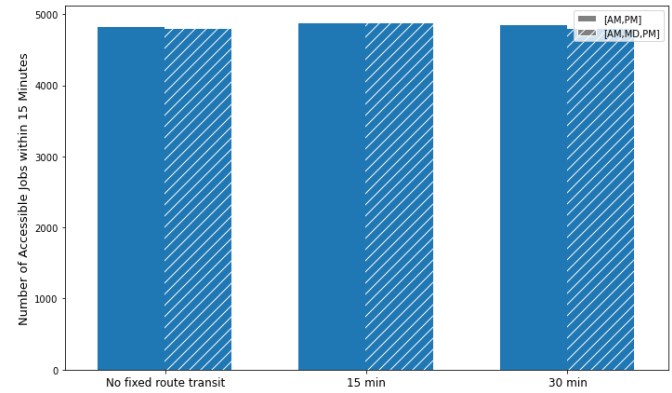
(b) 15-minute job accessibility

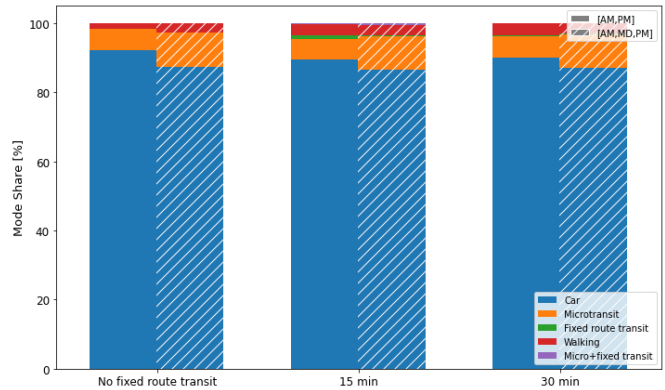
(c) mode share

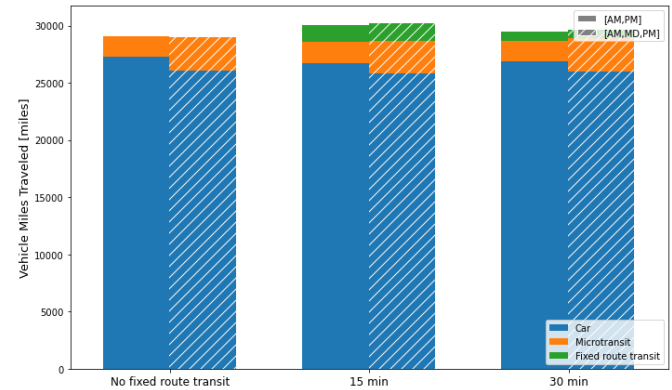
(d) vehicle miles traveled

Figure 17: Impacts of operating periods and transit frequencies for Lemon Grove network (virtual stop=75%, fleet size=10 vehicles)

## 6.3. Additional Model Outputs

This subsection illustrates several additional model outputs likely of interest to researchers and transportation planners.

### 6.3.1. Zonal 15-minute job accessibility

Figure 18 visualizes the number of jobs accessible within 15 minutes for travelers with a home origin from each zone in downtown San Diego. We compare FRT-only vs. an integrated fixed-flex PT system and 15-minute vs. 30-minute FRT headways in Figure 18.

Figure 18(a) and 18(c) display 15-minute job accessibility with 15-minute FRT headways, while Figure 18(b) and 18(d) display 15-minute job accessibility with 30-minute FRT headways. As the transit headway increases from 15 to 30 minutes (from left to right), the zones become less red, indicating that transit frequency positively correlates with zonal accessibility.

Figure 18(a) and Figure 18(b) display FRT-only scenarios while, Figure 18(c) and Figure 18(d) display integrated fixed-flex PT system scenarios. The results indicate that microtransit service increases zonal accessibility, albeit in a spatially diverse manner.

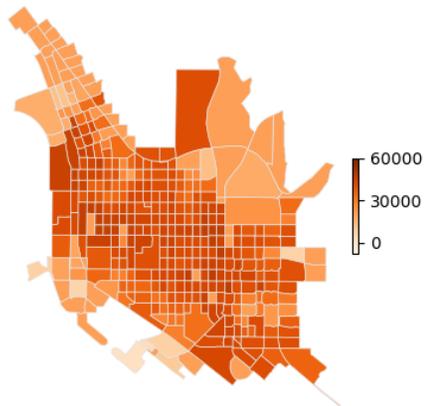

(a) Fixed transit only + 15-minute headway

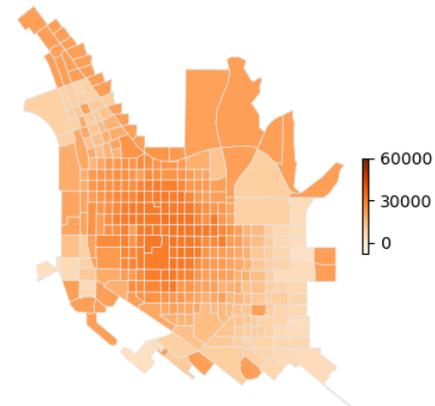

(b) Fixed transit only + 30-minute headway

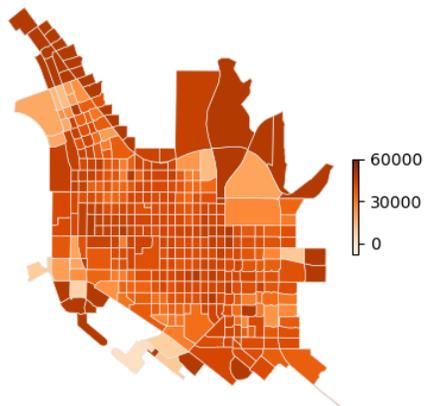

(c) Micro+fixed transit only + 15-minute headway

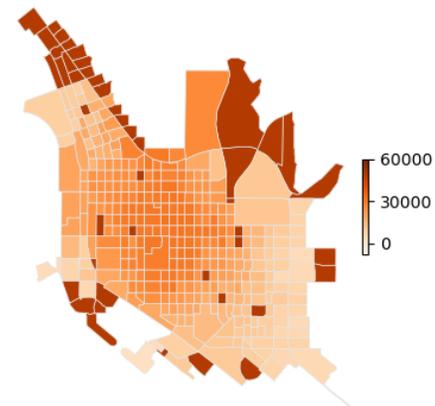

(d) Micro+fixed transit only + 30-minute headway

Figure 18: Zonal 15-minute job accessibility for downtown San Diego (fleet size: 10 veh, operating period: [AM, PM], virtual stop: 75%)

Figure 19 visualizes the number of jobs accessible within 15 minutes in each zone in Lemon Grove. The results are directionally the same as in downtown San Diego. However, in the case of Lemon Grove, the impact of microtransit service on accessibility is much greater than in San Diego. Microtransit service also has a bigger impact on accessibility than FRT line headways in Lemon Grove. These results are both expected and highly relevant to the design of integrated fixed-flex PT systems.

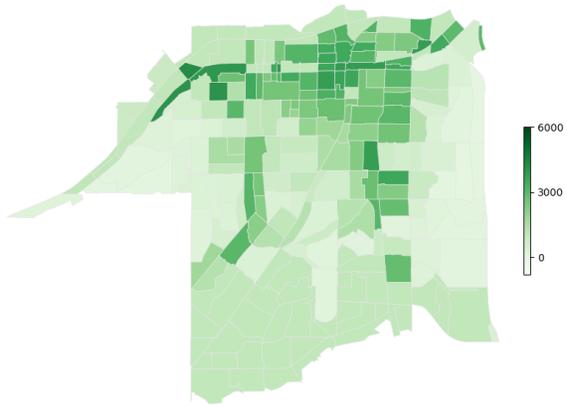
(a) Fixed transit only + 15-minute headway

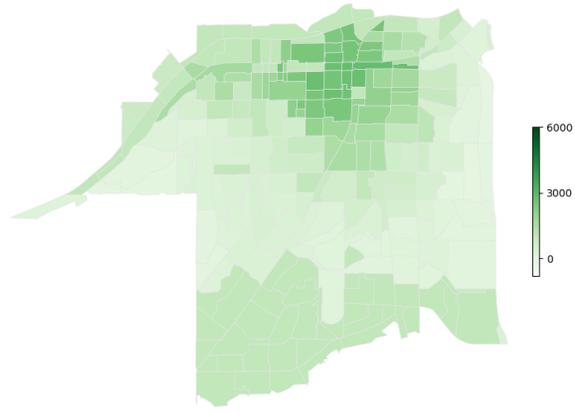
(b) Fixed transit only + 30-minute headway

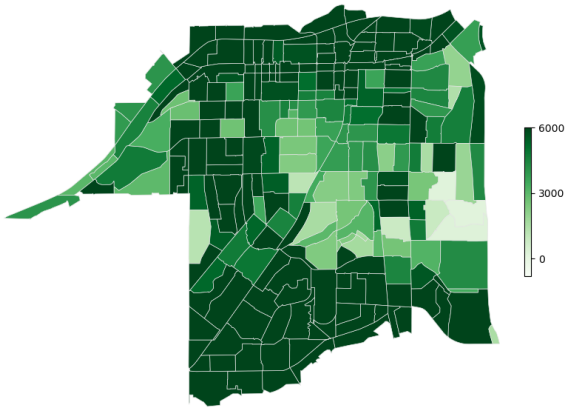
(c) Micro+fixed transit only + 15-minute headway

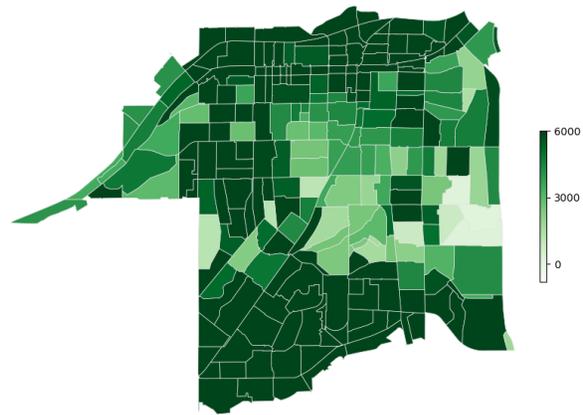
(d) Micro+fixed transit only + 30-minute headway

Figure 19: Zonal 15-minute job accessibility for Lemon Grove (fleet size: 10 veh, operating period: [AM, PM], virtual stop: 75%)

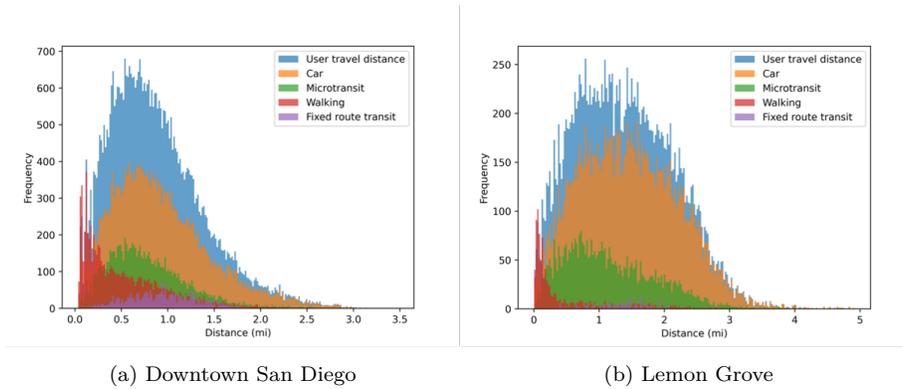

(a) Downtown San Diego  (b) Lemon Grove

Figure 20: Individual travel length distribution by mode (fleet size=10 veh, operating hour=[AM, MD, and PM]15h, headway=15min, virtual stop=100%)

### 6.3.2. Individual trip length distribution

Figure 20 shows the individual trip length distribution by mode for downtown San Diego (Figure 20(a)) and Lemon Grove (Figure 20(b)). The blue bins represent the distribution of the total travel distance of the travelers for a single trip. The orange, green, red, and purple bins represent the lengths of different trip legs for car, microtransit, walking, and FRT modes. For downtown San Diego, 95% of the trips have distances ranging from 0 to 1.84 miles, while for Lemon Grove, 95% of the trips have distances ranging from 0 to 2.67 miles.

For the walking mode, the trip length for downtown San Diego ranges from 0 to 0.5 miles, while for Lemon Grove, the walking trip length distribution is skewed towards 0. This shows that Lemon Grove has very few walking trips. For FRT, downtown San Diego has very few FRT trips (less than 7.5%) while Lemon Grove only has even smaller amount of FRT trips (less than 2%), due to its low FRT coverage.

Walking serves short distance trips (average trip length 0.53 miles in San Diego and 0.56 miles in Lemon Grove). Microtransit serves the short-to-middle distance trips (average trip length 0.79 miles in San Diego and 1.14 miles in Lemon Grove), while FRT serves middle-distance trips (average trip length 0.94 miles in San Diego and 1.22 miles in Lemon Grove). However, the auto mode serves trips of all distances.

### 6.3.3. Transit line usage

Figure 21 shows the transit line usage of downtown San Diego (21(a)) and Lemon Grove (21(b)) for Scenario 13 (headway = 15 minutes, virtual stop =75%, fleetsize = 10 veh, operating period = [AM, PM]). Transit link usage is first calculated as the number of travelers who use that transit link. Transit line usage is calculated by aggregating the transit link usage by route IDs. In terms of time of day, the midday (MD) period (orange bins) has a large share of link usage because microtransit does not operate during the MD period. In the downtown San Diego region, Route 4, Route 215, and Route 550 have a lower use of transit links, which means these routes do not serve many transit users. In the Lemon Grove region, Routes 520 and 4 have a lower use of transit links. These results show that the proposed model can estimate the FRT network usage after opening the microtransit service. So the proposed model could inform the FRT network design, for example, by pointing out the unproductive or low-usage lines. Transit agencies can use this information to re-design their transit network, such as removing some of the unproductive segments.

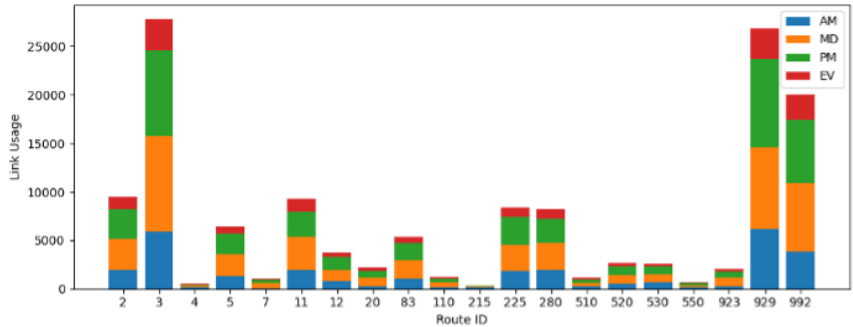

(a) Downtown San Diego

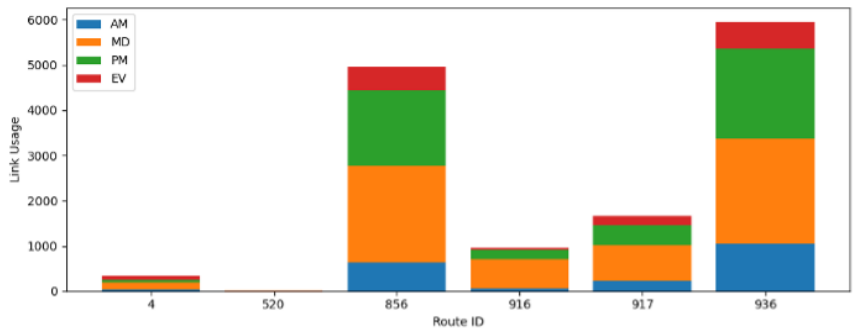

(b) Lemon Grove

Figure 21: Transit line usage classified by route ID (Scenario 13)

Figure 22 illustrates the transit line usage in downtown San Diego (Figure 22(a) and Figure 22(b)) and Lemon Grove (Figure 22(c) and Figure 22(d)). The result shows that the transit links in the center of downtown San Diego are heavily used, while the transit links in northwestern parts are rarely used. For Lemon Grove, northern part of the transit network has higher usage than the southern part of it. The transit links in downtown San Diego are thicker than those in Lemon Grove, because downtown San Diego has much higher ridership than Lemon Grove.

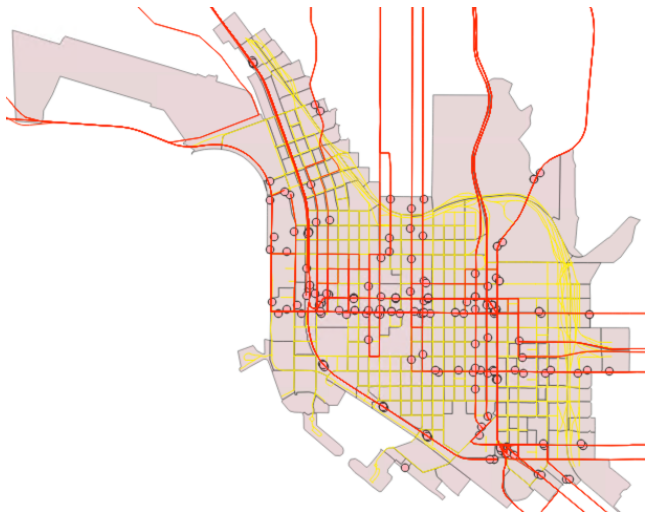

(a) Downtown San Diego Transit Network

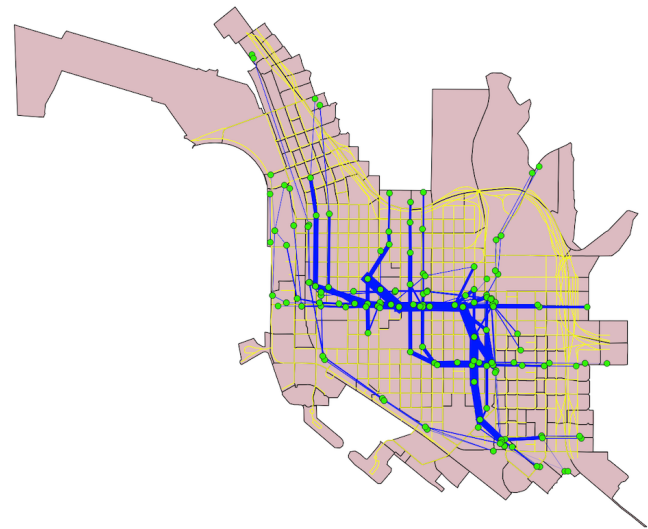

(b) Downtown San Diego Transit Usage

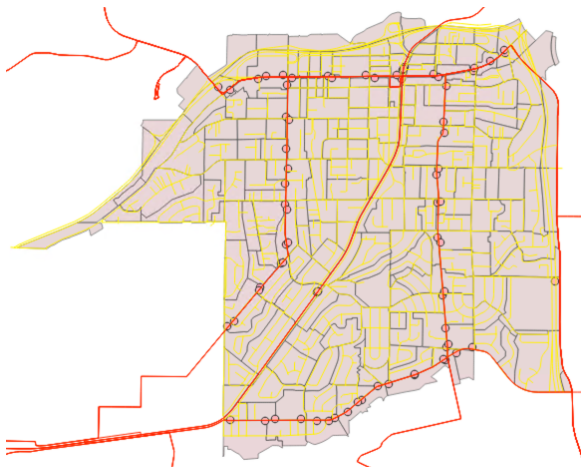

(c) Lemon Grove Transit Network

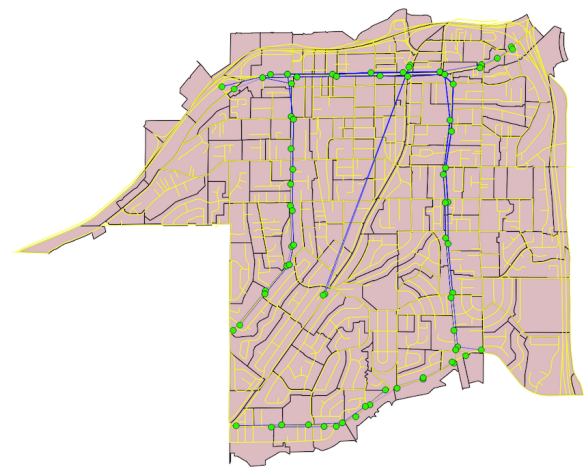

(d) Lemon Grove Transit Usage

Figure 22: Transit line usage (fleet size=10 veh, virtual stop=75%, operating period=[AM and PM])

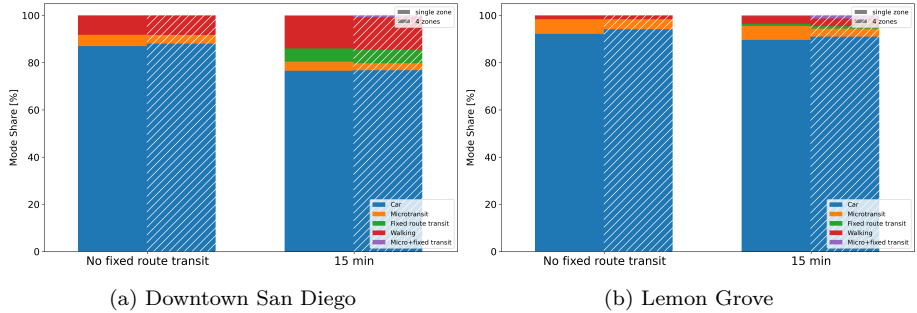

(a) Downtown San Diego

(b) Lemon Grove

Figure 23: Mode share before and after zonal partition (headway=15 min, fleet size=10 veh, virtual stop=75%, operating period=[AM and PM])

## 6.4. Impact of Zonal Partitioning

Figure 23 compares the results of mode share in the downtown San Diego (Figure 23(a)) and Lemon grove (Figure 23(b)) networks before and after zonal partitioning. After zonal partitioning, the car mode share and integrated fixed-flex PT system mode share increase in both downtown San Diego and Lemon Grove, while the microtransit mode share decreases.

Figure 24 compares total subsidy (revenue minus operating costs) in the downtown San Diego (Figure 24(a)) and Lemon Grove (Figure 24(b)) networks, respectively, before and after zonal partitioning. After partitioning, the total subsidy increases in both downtown San Diego and Lemon Grove, because microtransit revenue decreases, while FRT and microtransit costs are nearly the same.

Figure 25 compares the subsidy per passenger trip for each mode in downtown San Diego (Figure 25(a)) and Lemon Grove (Figure 25(b)) networks, respectively, before and after zonal partitioning. After partitioning, the total subsidy per passenger trip decreases in both downtown San Diego and Lemon Grove, because the number of FRT trips increases in both region.

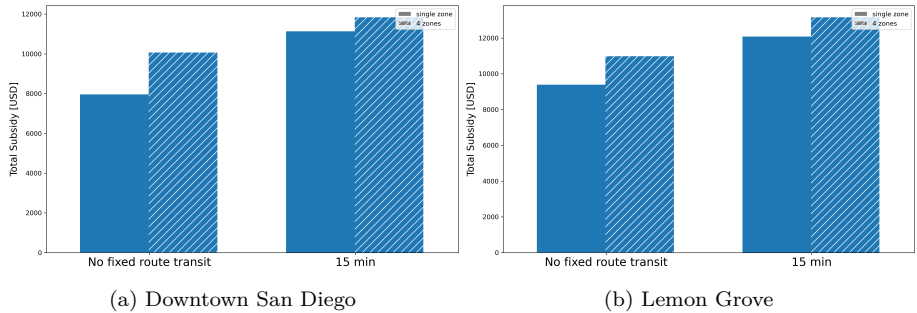

(a) Downtown San Diego

(b) Lemon Grove

Figure 24: Total subsidy before and after zonal partition (headway=15 min, fleet size=10 veh, virtual stop=75%, operating period=[AM and PM])

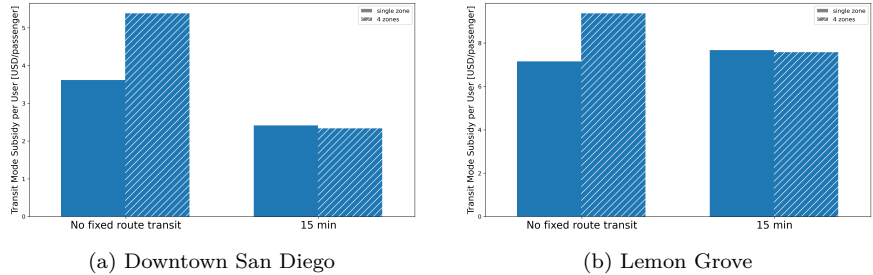

(a) Downtown San Diego  (b) Lemon Grove

Figure 25: Subsidy per passenger trip for each mode before and after zonal partition (headway=15 min, fleet size=10 veh, virtual stop=75%, operating period=[AM and PM])

In summary, partitioning the microtransit service region increases the integrated fixed-flex PT system mode share, but since microtransit revenue decreases, partitioning requires a larger subsidy.

## 7. Conclusion and Discussion

This paper proposes an integrated FRT and microtransit modeling framework to evaluate different system designs. The modeling framework includes a binary logit mode choice model for auto and composite transit (i.e., the integrated fixed-flex PT system); a multimodal supernetwork model with FRT, microtransit, and pedestrian infrastructure layers; and a microtransit fleet simulator. The modeling framework is simulation-based permitting us to evaluate a wide variety of integrate system designs and produce various performance metrics of interest to transit agencies. The model is agent-based to capture heterogeneity in traveler preferences and behaviors.

This paper names the research outcome as a modeling framework rather than a model because of its flexibility to incorporate different models within the framework. For example, the binary logit mode choice model in the demand module could be replaced by a multinomial logit mode choice model. The different layers in the supernetwork in the supply module can also be updated with different simulation results. For example, the constant FRT network could be replaced by a transit simulation, and the constant auto network could be replaced by a traffic simulation.

We use the proposed modeling framework to study different integrated fixed-flex PT system designs. We vary the following parameters: FRT frequencies, microtransit fleet size, service region structure, virtual stop coverage, and operating hours. We consider two case study regions in California, namely, downtown San Diego and Lemon Grove, a small city in San Diego County. We focus on four performance metrics: subsidy per transit user, number of jobs accessible within 15 minutes, mode share, and VMT.

Simulation results show that integrating microtransit with FRT will not necessarily increase transit ridership. Instead, it will decrease FRT ridership. However, integrating microtransit with FRT will reduce auto mode share and auto mode VMT, as well as increase the accessibility of the region. Small fleet size, high virtual stop coverage, and only operating during the peak period can lead to low subsidy per transit user. For areas with low transit coverage, such as Lemon Grove, opening microtransit service will lead to a high increase in accessibility. Microtransit could serve short- to medium-distance trips, while FRT could serve medium-distance trips. In contrast, auto mode can serve trips with all distances. The proposed modeling framework could also output the transit line usage, which provides information for transit agencies to redesign the transit network.

Future research efforts include developing a calibration algorithm to calibrate the individual coefficients for the choice model for the study region, and testing different microtransit fare structures. For example, if a traveler uses a microtransit service, he or she could get a free FRT service. Finally, the proposed modeling framework can also be incorporated with Bayesian optimization (Frazier, 2018) to find the optimal designs for the integrated FRT and microtransit system.

**Acknowledgements**

The first, second, and third authors received partial funding support for this study from the University of California Institute of Transportation Studies from the State of California for the California Resilient and Innovative Mobility Initiative (RIMI).

The first and second author received partial funding support from the National Science Foundation through CMMI 2125560, "SCC-IRG Track 1: Revamping Regional 803 Transportation Modeling and Planning to Address Unprecedented Community Needs during the Mobility Revolution."

The fourth author received partial support from the University of California Irvine Institute of Transportation Studies Transportation Research Immersion Program.

The authors also want to acknowledge Dr. Susie Pike and Dr. Dingtong Yang for their insights on microtransit and vehicle routing problems, respectively.

The authors are solely responsible for the contents of the manuscript.

Declaration of Conflict of Interests: None